%% file: main.tex
\documentclass[sigconf, nonacm]{acmart}  %

\keywords{Machine Learning for Security, Static Code Analysis, Code Representation, Web Application Security}

\ccsdesc[500]{Security and privacy~Web application security}
\ccsdesc[500]{Security and privacy~Software and application security}
\ccsdesc[500]{Security and privacy~Systems security}

\usepackage{enumitem}
\setlist[description]{
    itemsep=0pt, %
    parsep=0pt, %
    leftmargin=0pt %
}
\setlist[itemize]{
    itemsep=2pt, %
    parsep=2pt, %
    leftmargin=10pt %
}
\setlist[enumerate]{
    itemsep=2pt, %
    parsep=2pt, %
}

\usepackage{tcolorbox}
\tcbuselibrary{skins}
\newtcolorbox{answerbox}{
  colback=black!5!white,
  colframe=black!70!white,
  fonttitle=\bfseries,
  sharp corners,
  rounded corners,
  boxrule=0.3mm
}

\usepackage[utf8]{inputenc}
\usepackage{pgfplots}
\DeclareUnicodeCharacter{2212}{−}
\usepgfplotslibrary{groupplots,dateplot}
\usetikzlibrary{patterns,shapes.arrows, arrows.meta, positioning, arrows}
\pgfplotsset{compat=newest}

\usepackage{threeparttable}
\usepackage{booktabs}
\usepackage{tabularx}

\usepackage{mathrsfs} %

\AtBeginDocument{
\newcommand{\reqR}[1]{$\mathscr{R}$\textsubscript{#1}} %
\newcommand{\reqOne}{\hyperref[itm:requirement:1]{\reqR{1} - \textit{Conciseness}}}
\newcommand{\reqTwo}{\hyperref[itm:requirement:2]{\reqR{2} - \textit{Completeness}}}
\newcommand{\reqThree}{\hyperref[itm:requirement:3]{\reqR{3} - \textit{Simplicity}}}
\newcommand{\reqFour}{\hyperref[itm:requirement:4]{\reqR{4} - \textit{Stylistic Consistency}}}
\newcommand{\reqFive}{\hyperref[itm:requirement:5]{\reqR{5} - \textit{Computational Efficiency}}}
}

\usepackage{float} %

\usepackage{listings}

\usepackage{tikz}
\usetikzlibrary{shapes, positioning, intersections}
\usepackage{stmaryrd}

\usetikzlibrary{arrows, automata, positioning, decorations.pathreplacing, decorations.pathmorphing, shapes.gates.logic.US,shapes.gates.logic.IEC, shapes, intersections}

\usepackage{booktabs}
\usepackage{threeparttable}
\usepackage{multirow}
\usepackage{float}

\usepackage{amsmath}
\usepackage{wasysym} %
\usepackage[caption=false,font=footnotesize,labelfont=sf,textfont=sf]{subfig}
\usepackage{url}                %
\usepackage{hyperref}
\usepackage{cleveref}
\usepackage[disable]{todonotes} %

\input{setup}

\usepackage{orcidlink}

\begin{document}

\date{}

\title{
Trace Gadgets: Minimizing Code Context for Machine Learning-Based Vulnerability Prediction
}

\settopmatter{authorsperrow=4}

\newcommand{\affUzL}{%
  \affiliation{%
    \institution{ITS, University of Luebeck}
    \country{Luebeck, Germany}
  }%
}
\newcommand{\affUHH}{%
  \affiliation{%
    \institution{CHAI, University of Hamburg}
    \country{Hamburg, Germany}
  }%
}

\author{Felix Mächtle}
\orcid{0009-0009-2431-0322}
\affUzL

\author{Nils Loose}
\orcid{0009-0003-6243-1623}
\affUzL

\author{Tim Schulz}
\orcid{0009-0002-8409-7999}
\affUHH

\author{Florian Sieck}
\orcid{0000-0002-1501-0936}
\affUzL

\author{Jan-Niclas Serr}
\orcid{0009-0001-4624-8592}
\affUzL

\author{Ralf Moeller}
\orcid{0000-0002-1174-3323}
\affUHH

\author{Thomas Eisenbarth}
\orcid{0000-0003-1116-6973}
\affUzL

\renewcommand{\tablename}{Table}
\def\UrlBreaks{\do\/\do-}

\setlength{\abovedisplayskip}{5pt} %
\setlength{\belowdisplayskip}{5pt} %
\setlength{\abovedisplayshortskip}{5pt} %
\setlength{\belowdisplayshortskip}{5pt} %

\begin{abstract}

As the number of web applications and API endpoints exposed to the Internet continues to grow, so does the number of exploitable vulnerabilities.
Manually identifying such vulnerabilities is tedious. Meanwhile, static security scanners tend to produce many false positives.
While machine learning-based approaches are promising, they typically perform well only in scenarios where training and test data are closely related and the input length is short.
We demonstrate that excessively long contexts negatively affect the code comprehension capabilities of machine learning models, particularly smaller ones. Thus, 
a key challenge for ML-based vulnerability detection is providing suitable and concise code context.
This work introduces Trace Gadgets, a novel code representation that minimizes code context by removing irrelevant code.
Trace Gadgets precisely capture the statements that cover the path to the vulnerability. As input for ML models, Trace Gadgets provide a minimal but complete context, thereby improving the detection performance.
Moreover, we curate a large-scale dataset generated from real-world applications with manually curated labels to further improve the performance of ML-based vulnerability detectors.
Our results show that state-of-the-art machine learning models perform best when using Trace Gadgets compared to previous code representations, surpassing the detection capabilities of industry-standard static scanners such as GitHub's CodeQL by at least 4\% on a fully unseen dataset.
By applying our framework to real-world applications, we identify and report previously unknown vulnerabilities in widely deployed software. 
\end{abstract}

\maketitle

\input{sections/10-Introduction}

\input{sections/15-Background}
\input{sections/20-Code-Representation/0-Main}

\input{sections/30-Dataset}

\input{sections/40-Evaluation}

\input{sections/50-related-work}

\input{sections/60-Threats-to-Validity}

\section{Conclusion}
\label{sec:conclusion_and_future_work}
In summary, this research introduces \textit{Trace Gadgets}, a novel minimal code representation specifically designed for identifying injection vulnerabilities. 
A key motivation for developing such concise representations stems from our demonstration that increased code context length negatively impacts the code comprehension capabilities of machine learning models. %
This finding underscores the critical importance of minimizing input context for effective vulnerability detection. Trace Gadgets address this challenge by providing a minimal yet complete code context. 
Utilizing Trace Gadgets, we compiled a unique large-scale dataset, \textit{VulnDocker}, featuring real-world applications. %
To create labels for this dataset, we employed a leading industry scanner, namely FindSecBugs~\cite{find-sec-bugs}, and manually checked 10640 potentially vulnerable samples.
Using the VulnDocker dataset, we fine-tuned three advanced pre-trained machine learning models~\cite{ICSE2024_traced_ding,DBLP:ACL2022_Guo_Unixcoder,DBLP:Wang_CodeT5+}. Using those models and a commercial LLM, namely GPT-4o~\cite{gpt-4o}, we conducted a comparative analysis against both traditional static scanners and other ML-based frameworks and demonstrated a performance improvement over existing methods by $4{-}35\%$. In comparison with other code representations, namely slices (Code Gadgets) and functions, we show that Trace Gadgets achieve superior performance while requiring 28--34\% fewer tokens. 
Ultimately, Trace Gadgets and VulnDocker represent a significant advancement towards the practical application of ML for web-based vulnerability detection.

\begin{acks}
This work has been supported by the \textit{Federal Ministry of Research, Technology and Space of Germany (BMFTR)} through the \textit{PeT-HMR} and \textit{SILGENTAS} projects.
\end{acks}

\notefloi{Vllt. noch mehr den Bytecode Ansatz}

\bibliographystyle{plain}
\bibliography{
    references.bib,
    References/Rule-Based-Vulnerability-Detection.bib,
    References/Similarity-Based-Vulnerability-Detection.bib,
    References/Pattern-Based-Vulnerability-Detection.bib,
    References/Misc.bib
}

\appendix
\input{sections/70-Appendix}

\end{document}

%% file: setup.tex
\definecolor{highlightcolor}{RGB}{139, 195, 74}
\definecolor{highlightcolor2}{RGB}{0, 151, 167}

\lstdefinestyle{uzl-java}{
    frame=tb,
    language=Java, 
    basicstyle=\ttfamily\scriptsize, %
    stringstyle={\color{uzlpal4a}}, 
    keywordstyle={\color{uzlmain}}, 
    commentstyle={\color{uzlgrey}}, 
    showspaces=false, 
    showtabs=false, 
    showstringspaces=false, 
    breaklines=true, 
    breakatwhitespace=false,
    firstnumber=1, 
    numbers=left, 
    numberstyle=\tiny\ttfamily\color{uzlgrey}, 
    numberblanklines=false, 
    numbersep=.5em,
    xleftmargin=1.2em, 
    framexleftmargin=1em, 
}

\definecolor{uzlpal4a}{rgb}{0.0, 0.6, 0.0} %
\definecolor{uzlmain}{rgb}{0.0, 0.0, 0.6}  %
\definecolor{uzlgrey}{gray}{0.5}           %

\newcommand{\bheading}[1]{{{\textbf{#1.}}}}

\makeatletter
\def\namedlabel#1#2{\begingroup
  #2%
  \def\@currentlabel{#2}%
  \phantomsection\label{#1}\endgroup
}
\makeatother

\newcommand{\notefelixi}[1]{\todo[color=green!30,inline]{FM: #1}}

\newcommand{\notefloi}[1]{\todo[color=blue!20,inline]{FS: #1}}

\newcommand{\TGs}{\textit{TGs}}
\newcommand{\TG}{\textit{TG}}
\newcommand{\TraceGadget}{Trace Gadget}

%% file: sections/10-Introduction.tex
\section{Introduction}
\label{sec:introduction}

Software vulnerabilities pose a significant threat to organizations in all sectors. Systems with public interfaces are especially vulnerable to attacks. 
If these systems are hosted in critical environments such as hospitals, banks or transportation systems, their exploitation has a devastating impact \cite{wannacry,malware_hospitals}. 
\notefloi{Why ransomware examples?}
The rising number of web services and internet facing API endpoints leads to an immense attack surface where exposed endpoints introduce a significant security risk \cite{DBLP:conf/secdev/Chen19}.

Yet, traditional security methods, such as penetration testing, require labor-intensive manual inspection by human experts, making penetration testing approaches costly and time-consuming. 
Considering the shortage of experts, the industry often turns to automated scanners \cite{CodeQL, Semgrep, find-sec-bugs} as a cost-effective alternative or addition.
However, these scanners, which rely solely on rule-based systems without deep code understanding, fall short compared to human experts \cite{humans_are_better_than_static_scanners}. 
While they report high true positive (TP) rates, they are often accompanied by high false positive (FP) rates, leading to numerous false incidents. %
\par   
With recent successes 
in the area of natural language processing (NLP) \cite{first_transformer, DBLP:journals/corr/abs-2302-09419}, the capabilities of machine learning (ML) models to reason about code have advanced. 
State-of-the-art models such as \textit{UniXcoder}~\cite{DBLP:ACL2022_Guo_Unixcoder}, \textit{Traced}~\cite{ICSE2024_traced_ding}, \textit{CodeT5+}~\cite{DBLP:Wang_CodeT5+} or \textit{GPT-4o}~\cite{gpt-4o} report promising results in code understanding~\cite{ICSE2024_traced_ding, DBLP:Wang_CodeT5+}.
These advances have led to an active area of research focused on leveraging ML techniques to detect vulnerabilities in existing software~\cite{DBLP:ICML2021_Peng_OSCAR, ReGVD, VulDeePecker, DeepWukong, sentence-encodings, DBLP:Achilles, DBLP:conf/compsac/Partenza, DBLP:conf/qrs/Mamede/2022}.
The goal is to be able to detect vulnerabilities in real-world datasets without the need to manually define vulnerability patterns. %

\input{figures/Slicing-Example}

However, machine learning techniques struggle with long inputs and excessive noise. Although modern commercial LLMs such as GPT-4o~\cite{gpt-4o} or Claude Sonnet~\cite{claude-sonnet} can theoretically handle up to 128K+ tokens, their performance degrades as the input length increases (\Cref{sec:eval:llm-context-length}). While most research has focused on modifying the input or exploring different models, little work has been devoted to minimizing the context length for vulnerability prediction~\cite{DBLP:journals/corr/abs-2403-10646}. 
Most approaches represent the input using function-level representations~\cite{DBLP:conf/msr/FuT22/LineVul, DBLP:conf/icse/SteenhoekGL24, DBLP:conf/nips/ZhouLSD019/Devign, DBLP:Achilles, sentence-encodings, DBLP:conf/compsac/Partenza, DBLP:conf/nips/ZhouLSD019/Devign} or program slicing~\cite{DeepWukong, VulDeePecker}.
However, these representations often include irrelevant code, which introduces noise and impairs predictive performance~\cite{DBLP:journals/corr/abs-2505-07897/long-code-benchmark, DBLP:journals/corr/abs-2503-04359/long-code-benchmark}. Both representation also lack stylistic consistency, thus preventing the model from focusing solely on the substance rather than the form of the code and increasing its cognitive load~\cite{DBLP:Code_Representation_must_be_consistent, DBLP:conf/uss/RisseB24, DBLP:conf/sp/UllahHPPCS24}.
Furthermore, function-level representations omit essential contextual information, which limits their effectiveness~\cite{DBLP:conf/sigsoft/WuLXLS023, DBLP:conf/uss/MirskyMBYPDML23/VulChecker}.

In this work, we argue that effective ML-based vulnerability detection requires a task-specific code representation that is both concise and semantics-preserving. 
We focus on injection vulnerabilities, a major class of web vulnerabilities~\cite{owasp-top-10} where a user-controlled value reaches a vulnerable statement without proper sanitization. For those vulnerabilities, we construct the minimal context that still captures the relevant data and control flow.

\bheading{Trace Gadgets and framework overview} 
To address these challenges, we introduce \emph{Trace Gadgets} (\TGs{}), a novel code representation that combines static tracing with slicing around potentially vulnerable statements. 
Similar to \emph{Code Gadgets}~\cite{VulDeePecker}, the code is sliced with respect to a potentially vulnerable statement. However, before slicing we statically trace the code to remove irrelevant statements (\Cref{sec:code-representation}). 
This yields short, semantically correct  snippets that significantly reduce input length while retaining the essential control and data flow for the detection of injection vulnerabilities.

The source code for software applications
is not always freely available and users might only have access to
binaries or bytecode. Since Trace Gadgets are defined purely by
control and data flow semantics, any representation that exposes
them can be transformed into \TGs{}.
In our prototype, we instantiate this idea on JVM bytecode: the JVM ecosystem (e.g. Java, Kotlin, Scala) is widely used for web applications~\cite{statista-java-most-used-languages,tiobe-index}, and its well-defined bytecode format facilitates static analysis. 
All our experiments therefore operate on JVM bytecode, enabling us to analyze both closed-source proprietary web applications and open source projects.

\bheading{Research questions}
To verify and evaluate whether \TGs{} are a suitable code representation for vulnerability detection and satisfy the requirements posed above, we design the following research questions.
\label{sec:research-questions}
\begin{enumerate}[leftmargin=25pt, itemsep=0pt]
    \item[\namedlabel{item:rq0}{\textbf{RQ1}}] How does the length of input context affect the code comprehension performance of modern LLMs? (\Cref{sec:eval:llm-context-length})
    
    \item[\namedlabel{item:rq1}{\textbf{RQ2}}] What is the efficiency and effectiveness of our Trace Gadget generation engine? (\Cref{sec:eval:tracegadetgeneration})

    \item[\namedlabel{item:rq2}{\textbf{RQ3}}] 
    Can state-of-the-art ML models be effectively combined with Trace Gadgets for vulnerability classification on unseen datasets? (\Cref{rq:rq2})

    \item[\namedlabel{item:rq5}{\textbf{RQ4}}] 
    Can the minimal context of Trace Gadgets enhance the overall vulnerability detection performance of state-of-the-art? (\Cref{sec:eval:detection-performance-comparison}) 

    \item[\namedlabel{item:rq4}{\textbf{RQ5}}] Do Trace Gadgets fulfill the proposed code representation requirements for vulnerability detection? (\Cref{sec:rq:answer:do_trace_gadgets_fulfill_requirements})    
\end{enumerate}

\bheading{Dataset and real-world evaluation} 
To evaluate the proposed research questions, we  
require large and labeled datasets. 
For JVM web services, existing datasets are either large but synthetic~\cite{nsa-juliet-dataset, owasp-java}, raising concerns about their realism~\cite{DBLP:conf/compsac/Partenza}, or small and application-specific~\cite{DBLP:conf/msr/BuiSF22/vul4j}, limiting their usefulness for training general models. Therefore, we construct a new dataset that overcomes the limitations. Because we work on bytecode instead of source code, we can analyze compiled, production-ready programs. 
We therefore collect JVM web applications from Docker Hub, extract Trace Gadgets from the contained Jar files, and label them with a combination of static analysis and partial manual verification, resulting in a large-scale, real-world dataset that we call \emph{VulnDocker}.

\bheading{Framework and practical impact} 
Building on Trace Gadgets and VulnDocker, we develop a vulnerability detection framework, depicted in \Cref{fig:paper-overview}, that integrates \TGs{} with state-of-the-art code models and LLMs. 
We use this framework to train and evaluate multiple models and code representations, as well as to compare against industrial static scanners~\cite{CodeQL, Semgrep, find-sec-bugs}. 
Our evaluation shows that \TGs{} not only improve detection performance compared to alternative code representations, but also outperform established industrial static analysis tools. We open source this framework on GitHub: \textit{\href{https://github.com/UzL-ITS/Trace-Gadgets}{https://github.com/UzL-ITS/Trace-Gadgets}}.
In addition, %
our framework with \TGs{} uncovered two previously unknown vulnerabilities in widely deployed web applications (Atlassian Bamboo and Geoserver), both of which have been responsibly disclosed by us and fixed by the vendors~\cite{Bugcrowd-Finding1, patch-geoserver}.

\input{figures/Paper-Overview}
\bheading{Contributions}
To summarize, our contributions are:
\begin{itemize}[leftmargin=10pt]

    \item We demonstrate that, for security tasks, modern Large Language Models, despite theoretically supporting large context sizes, exhibit reduced code comprehension performance for longer inputs.

    \item We introduce \textit{Trace Gadgets}, a novel code representation that preserves the original program semantics while reducing code length by 28--34\% compared to existing code representation. Using \TGs{} as input reduces the False Positive Rate by 29–38\% in related methods.\notefloi{Same remark as before}

    \item We create a large-scale labeled dataset of Trace Gadgets based on real-world JVM applications retrieved from DockerHub, consisting of 32886 deduplicated samples with manually curated labels.

\end{itemize}

%% file: figures/Slicing-Example.tex
\begin{figure}[t!]

\begin{lstlisting}[style=uzl-java, escapechar=|, caption={Motivating Example: A Trace Gadget generated from the OWASP Benchmark test case \textit{BenchmarkTest01314}. The original test case spans multiple classes, conditions and functions, but is distilled into a precise, single-function representation via code inlining and the removal of redundant computations, thereby isolating the sink-relevant functionality.\\}, label={lst:slice-example}]
void TG(HttpServletRequest var0,
           HttpServletResponse var1){
  String var2 = var0.getParameter(
                "BenchmarkTest01314"); // Source
  if (var2 == null) var2 = "";
  var2 = "INSERT INTO users (username, password) VALUES ('foo','" + var2 + "')";
  Statement var3 = DBHelper.getSqlStatement();
  var3.executeUpdate(var2); // Sink
}

\end{lstlisting}
\end{figure}

%% file: figures/Paper-Overview.tex
\begin{figure}[t]
    \centering
\includegraphics[width=\columnwidth]{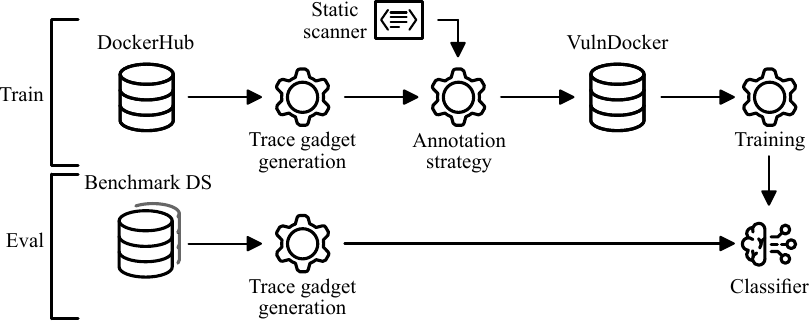}

    \caption{Systematic overview of the training and evaluation phase. %
    }
    \label{fig:paper-overview}
\end{figure}

%% file: sections/15-Background.tex
\section{Background}

\notefloi{True positive rate, FPR and TPR=FPR=1 is guessing}

\subsection{Program Slicing}
Program slicing~\cite{program_slicing} is a technique used to reduce a program to only those statements necessary for a particular computation, known as the \emph{slicing criterion}. This reduction is achieved by constructing a Program Dependence Graph (PDG), which integrates both the Control Flow Graph (CFG) and the Data Flow Graph (DFG) to represent the dependencies between program statements. By performing a reachability analysis from a given node in the PDG, either forward or backward, it is possible to determine which statements influence or are influenced by the slicing criterion. %

\subsection{Code Gadgets}
Li \emph{et al.} introduced \textit{Code Gadgets} as a representation for vulnerability detection~\cite{VulDeePecker}. In their approach, program slicing is performed with respect to a list of potentially vulnerable statements. The resulting slices, composed of lines of code, are then assembled into single code snippets suitable for machine learning. Since these code snippets contain all relevant statements from the PDG, they effectively represent all possible executions that could lead to the execution of a particular sink.

\subsection{Program Traces}
\label{sec:background:program-traces}

Program tracing is a technique used to monitor the execution of a program by recording the sequence of statements as they are executed~\cite{DBLP:conf/pldi/0022S20/LiGer/Traces}. A \emph{program trace} describes a particular execution path through a program $\mathcal{P}$ consisting of statements $s_0, s_1, \dots, s_n$, where each $s_i$ belongs to a set of statements $\mathcal{S}$.

To model the entire program $\mathcal{P}$, it is necessary to consider the set of all possible execution traces, accounting for every potential path the program might take during execution. This set of traces $\mathcal{T}$ collectively represents the program's behavior and can be expressed as:
\[
\mathcal{P} = \bigcup_{t \in \mathcal{T}} t
\]
where each trace $t \in \mathcal{T}$ is a sequence of statements from $\mathcal{S}$ corresponding to a possible execution path. 
\Cref{listing:source-code-example} shows a simple example function whose traces would be $ t_1 = [1,2,3,4,8,9] \text{ and } \: t_2 = [1,2,3,6,8,9]$.

\subsection{Injection vulnerabilities}
\label{sec:background:injection-vulnerabilites}

Injection vulnerabilities are one of the most significant and prevalent security risks in web applications~\cite{owasp-top-10}. These occur when attacker-controlled input enters the program at a source statement ($s_k \in \mathcal{S}_{\text{source}}$), such as line 4 in~\Cref{fig:listing-of-3-code-snippets-from-source-to-trace-gadgets}, and propagates to a sink ($s_l \in \mathcal{S}_{\text{sink}}$) without proper sanitization~\cite{DBLP:conf/sp/YamaguchiMGR15/TaintStyleVulnsDefinition}.

\subsection{Java Vulnerability Datasets}
\bheading{NIST Juliet Java 1.3} The National Institute of Standards and Technology (NIST) provides several vulnerability datasets, including the \textit{Juliet Java 1.3} test suite \cite{nsa-juliet-dataset}. 
The Juliet dataset contains various small vulnerability samples categorized by their Common Weakness Enumeration (CWE). Each vulnerability is documented, classified and labelled. However, the dataset's synthetic nature and small toy examples limit the suitability for training ML models for detecting vulnerabilities in real-world applications.
Additionally, Partenza \emph{et al.}~\cite{DBLP:conf/compsac/Partenza} observed that the dataset is biased due to different lengths of benign and vulnerable examples. This bias results in artificial discriminating features for ML models.

\bheading{OWASP Benchmark}
The Open Web Application Security Project~(OWASP) provides a vulnerability dataset~\cite{owasp-java}, focusing on the top ten web application security risks~\cite{owasp-top-10}. 
Similar to the Juliet test suite, each test case in this dataset is an endpoint annotated with labels indicating its vulnerability status. In contrast to Juliet, the OWASP Benchmark presents a greater number of execution paths per endpoint, thus offering a more challenging environment for testing security tools.

%% file: sections/20-Code-Representation/0-Main.tex
\section{Code representation}

\begin{table}[t]
    \centering
    \caption{Comparison of different code representations}
    \label{tab:code-representation:comparison-table}
\begin{tabularx}{.95\columnwidth}{Xccccc}
                      & \reqR{1}                         & \reqR{2}                     & \reqR{3}                     & \reqR{4}                     & \reqR{5}                     \\
Function-level~\cite{DBLP:conf/msr/FuT22/LineVul, DBLP:conf/icse/SteenhoekGL24, DBLP:conf/nips/ZhouLSD019/Devign, DBLP:Achilles, sentence-encodings, DBLP:conf/compsac/Partenza, DBLP:conf/nips/ZhouLSD019/Devign}              & \Circle     & \Circle & \CIRCLE & \Circle & \CIRCLE \\
Slices (Code Gadgets)~\cite{VulDeePecker, DeepWukong, DBLP:conf/eurosp/KrakerVH23/Glice, DBLP:conf/uss/MirskyMBYPDML23/VulChecker} & \LEFTcircle & \CIRCLE & \Circle & \Circle & \CIRCLE \\
Execution Traces~\cite{DBLP:conf/pldi/0022S20/LiGer/Traces, DBLP:conf/icse/WangTTWLFX024/Concoction}         & \Circle     & \CIRCLE & \CIRCLE & \Circle & \LEFTcircle \\
Trace Gadgets         & \CIRCLE     & \CIRCLE & \CIRCLE & \CIRCLE & \LEFTcircle
\end{tabularx}
    
\end{table}

\input{figures/Tracing-Example-3-snippets}
\label{sec:code-representation}
In ML-based vulnerability detection, the code representation plays a vital role. It %
is the basis for the subsequent ML processes. Therefore, the quality and effectiveness of the code representation has a significant impact on the overall performance of the vulnerability detection system. %
Thus, a key challenge is to determine the optimal code representation~\cite{DBLP:conf/internetware/ZhangZYWJ23/ComparingDifferentCodeRepresentations}. This involves carefully considering what information should be included. The representation must be precise, capture relevant code patterns and logic necessary for identifying vulnerabilities while avoiding redundant information that could dilute the model's focus and efficiency.
To identify a suitable representation $\mathcal{R}$, we collect and extend criteria from related works:

\begin{itemize}[leftmargin=17pt]
\item[\label{itm:requirement:1}\reqR{1}] \textit{Conciseness} - %
$\mathcal{R}$ should be compact, mimicking the human ability to spot vulnerabilities in shorter code snippets compared to longer ones~\cite{DBLP:journals/corr/abs-2505-07897/long-code-benchmark, DBLP:journals/corr/abs-2503-04359/long-code-benchmark}. In \Cref{sec:eval:llm-context-length} we describe and evaluate an experiment that emphasizes the importance of \reqR{1} for security related tasks. %
\item[\label{itm:requirement:2}\reqR{2}] \textit{Completeness} - It is crucial that $\mathcal{R}$ encompasses all vital statements that contribute to pinpointing vulnerabilities. This ensures that the model has all the information necessary for precise and accurate vulnerability detection~\cite{DBLP:conf/sigsoft/WuLXLS023, DBLP:conf/uss/MirskyMBYPDML23/VulChecker}. %
\item[\label{itm:requirement:3}\reqR{3}] \textit{Simplicity} - The representation should aim for simplicity, limiting its structural elements, such as the number of functions or classes, to as few elements as necessary for comprehensibility~\cite{DBLP:journals/corr/abs-2311-16169/OWASP-LLM}. %
\item[\label{itm:requirement:4}\reqR{4}] \textit{Stylistic Consistency} - Maintaining a consistent programming style in $\mathcal{R}$ is crucial \cite{DBLP:Code_Representation_must_be_consistent, DBLP:conf/uss/RisseB24, DBLP:conf/sp/UllahHPPCS24}. This uniformity helps the model to focus on the substance rather than the form of the code, thereby reducing the cognitive load on the model and potentially enhancing its effectiveness.
\item[\label{itm:requirement:5}\reqR{5}] \textit{Computational Efficiency} - While ensuring the quality of the representation remains uncompromised, it is imperative that generating $\mathcal{R}$ is computationally efficient. 
\end{itemize}

\reqR{1}--\reqR{4} were collected from related work. Additionally, we included \reqR{5} to ensure the practical feasibility of integrating a representation into real-world pipelines, where large-scale code bases and time-sensitive analyses require fast preprocessing.
A categorization of existing granularities is shown in~\Cref{tab:code-representation:comparison-table} and explained in the following paragraphs:

\bheading{Function-level} Often, a function-level granularity is chosen as granularity~\cite{sentence-encodings, DBLP:Achilles, ReGVD, DBLP:conf/compsac/Partenza}, where each single function is individually analyzed without further context. 
This granularity is computationally efficient, thereby fulfilling \reqFive, and simple due to its limited scope, addressing \reqThree.
However, it falls short in several areas. By analyzing an entire function, it includes extraneous statements that are irrelevant to the vulnerability being targeted, thus violating \reqOne. Additionally, function-level granularity fails to capture relevant statements if a vulnerability spans multiple functions or classes, which compromises \reqTwo. Lastly, without further preprocessing, the code representation still retains the original programming style and variable naming conventions, leading to a violation of \reqFour.

\par
\par
\bheading{Slices (Code Gadgets)} 
When additional information is provided, such as a \textit{sink} statement $s_{sink}$, program slicing \cite{program_slicing} can be used to reduce the number of statements to those that affect the sink. This technique ensures that only the relevant statements are retained, thereby supporting \reqOne. Nevertheless, the approach does not entirely eliminate all execution paths within a single slice, which means it only partially addresses this requirement.
Nonetheless, program slicing remains computationally efficient to perform, thereby fulfilling \reqFive. When combined with an inter-procedural analysis, all relevant statements are included, thus meeting \reqTwo.
However, this broader scope can introduce multiple functions or classes, which compromises simplicity, thereby violating \reqThree.
Additionally, similar to function-level granularity, without further preprocessing, Code Gadgets still reflect the original programming style, thereby failing to satisfy \reqFour. 

\textbf{\bheading{Execution Traces}} Capturing the execution of a program captures every line of code that is executed, ensuring that all relevant statements are included. This satisfies \reqTwo. In addition, since the trace is derived from a single execution run, \reqThree{} is inherently fulfilled.
However, execution traces may contain numerous auxiliary computations that are unrelated to the vulnerability in question, thus failing to meet \reqOne. Furthermore, because the recorded source lines are aggregated without any enforced uniformity, does not fulfill \reqFour. Finally, obtaining such traces requires actual program execution, and often requires steering execution toward specific sinks using resource-intensive techniques such as symbolic execution, thereby compromising \reqFive.

\subsection{Trace Gadgets}
\label{subsec:trace-gadget}
\input{sections/20-Code-Representation/Trace-Gadget-Neu}

\subsection{Implementation}
\label{subsec:toolchain}

\input{sections/20-Code-Representation/2-Our-Toolchain}

%% file: figures/Tracing-Example-3-snippets.tex
\begin{figure*}[t]
    \centering
    \begin{minipage}{0.32\textwidth}
    \begin{lstlisting}[style=uzl-java,escapechar=|, caption={Source Program}, label=listing:source-code-example]
void doGet(HttpServletRequest rq){
  String p;  
  if (rq.getParameter("A")!=null){
    p = rq.getParameter("A");
  } else {
    p = "DEFAULT"; 
  }
  Log.debug("Database update");
  DB.executeUpdate(p); // Sink
}
        \end{lstlisting}
    \end{minipage}\hfill
    \begin{minipage}{0.32\textwidth}
    \begin{lstlisting}[style=uzl-java,escapechar=|, caption={Trace Gadget 1}]
void doGet(HttpServletRequest v1){
  String v2;  
  if (v1.getParameter("A")!=null){
    v2 = v1.getParameter("A");
  }

  DB.executeUpdate(v2); // Sink
}
        \end{lstlisting}
    \end{minipage}\hfill
    \begin{minipage}{0.32\textwidth}
    \begin{lstlisting}[style=uzl-java,escapechar=|, caption={Trace Gadget 2}]
void doGet(HttpServletRequest v1){
  String v2;

  if (v1.getParameter("A")==null){
    v2 = "DEFAULT"; 
  }
  
  DB.executeUpdate(v2); // Sink
}
        \end{lstlisting}
    \end{minipage}

\caption{Trace Gadget Generation: The logging statement is removed in both gadgets as it does not influence the value that flows into the sink. The \texttt{IF}-statement is split into \texttt{then} and \texttt{else} branches, each represented by one of the two Trace Gadgets.}
    \label{fig:listing-of-3-code-snippets-from-source-to-trace-gadgets}
\end{figure*}

%% file: sections/20-Code-Representation/Trace-Gadget-Neu.tex
Our novel code representation is developed, aligned with the previously established criteria, by building upon the concept of Code Gadgets, i.e., inter-procedural program slicing~\cite{program_slicing} used for source to sink vulnerabilities. 
Therefore, static program tracing is combined with Code Gadgets to generate \textit{Trace Gadgets}~(\textit{TGs}).
By slicing a trace, the number of statements is further reduced while retaining all relevant ones. Furthermore, the code from multiple classes and functions is merged into a single function to simplify further analysis.
An example of such a transformation is shown in~\Cref{fig:listing-of-3-code-snippets-from-source-to-trace-gadgets}.
Instead of a single Code Gadget, this approach results in multiple \TGs{} (two in the referenced example). 
Each \TG{} only contains the statements relevant to a particular sink %
in a single function. %

\bheading{Formalization} \label{sec:one-sided-branches}
Formally, the concept of tracing, as introduced in \Cref{sec:background:program-traces}, is used as a foundation. 
However, using the traditional concept of tracing, the number of traces is exponential in the number of all branches as every trace only contains a single control flow path.
However, experimental evaluation on web applications have shown that many conditional statements contain only a single branch while the other branch is effectively empty ($\emptyset$). Such conditions are often used to enforce a boundary. 
Hence, to avoid exponential behavior for one-sided branches, we 
include elements that represent one-sided conditional statements, into the analytical model of traces. 
Formally, we denote an \textit{if-then-else} construct as $\left(c_j, s_{j + 1}^{\mathcal{T}}, s_{j + 1}^{\mathcal{F}} \right)$, where $c_j$ is the condition, $s_{j + 1}^{\mathcal{T}}$ is the first instruction of the \textit{then} branch, and $s_{j + 1}^{\mathcal{F}}$ is the first instruction of the \textit{else} branch. 
Thus, we define the extended model $\mathcal{S}'$ as:
\[\mathcal{S}' = \mathcal{S} \cup \left \{ \left (c_j, s_{j + 1}^{\mathcal{T}}, \emptyset \right) \right \} \]

\input{figures/Toolchain-Overview}

As explained in \Cref{sec:background:injection-vulnerabilites}, all source to sink vulnerabilities, such as injections, contain a source statement ($s'_k \in \mathcal{S'}_{source}\subset \mathcal{S'}$) later followed by a sink statement ($s'_l \in \mathcal{S'}_{sink}\subset \mathcal{S'}$). 
Hence, only traces containing these two in the correct order are of interest. 
Notably, we adapt the use of the 'in' operator ($\in$) to check for the presence of a statement $s$ in a trace $t$ which is a sequence of statements:
\[ \exists s'_k, s'_l \in t_i: s'_k \in \mathcal{S'}_{source} \land s'_l \in \mathcal{S'}_{sink} \land k < l \]
Additionally, slicing is used to further remove statements within a trace that are irrelevant to the intended sink $s'_l$. 
Combining the previously mentioned modification, Trace Gadgets, by design, no longer model the entire program $\mathcal{P}$. Let $\mathcal{TG}$ be the set of all Trace Gadgets. We define:

\begin{equation*}
\begin{split}
t \in \mathcal{TG} \iff \:\:
      &s'_k \in t \land s'_l \in t \land k < l  \\
\land &\forall s'_i \in t: s'_i \text{ is relevant for } s'_l \\ 
\end{split}
\end{equation*}

While this work focuses on JVM-based analysis of injection vulnerabilities in the next sections, the representation can be applied to any programming language and vulnerabilities with a source to sink behavior.

%% file: figures/Toolchain-Overview.tex
\begin{figure*}[ht!]
    \centering
\includegraphics[width=0.9\textwidth]{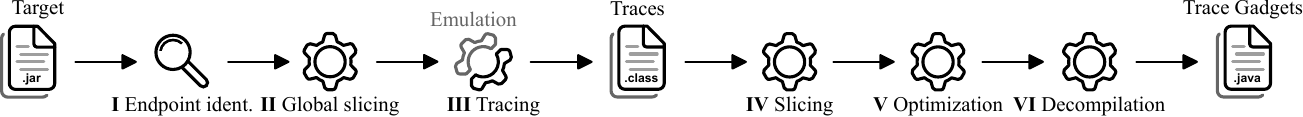}
    \caption{Systematic overview of the code preparation steps. The Roman numerals denote the paragraphs in \Cref{subsec:toolchain}.}
    \label{fig:system-overview}
\end{figure*}

%% file: sections/20-Code-Representation/2-Our-Toolchain.tex
This section introduces the engine developed to generate Trace Gadgets for JVM-based web applications. %
\Cref{fig:system-overview} shows an overview of our pipeline. A target Java archive (\texttt{Jar}) file is first analyzed to identify all possible endpoints (\textbf{I}). For each endpoint, Global Slicing identifies all functions that occur between a single source and a single sink (\textbf{II}). 
Each endpoint is then used as an entry point, along with its associated functions from \textbf{II}, to generate traces using a custom static emulation engine (\textbf{III}). Each trace is sliced (\textbf{IV}) before being optimized (\textbf{V}). Finally, the optimized \TG{} is decompiled.

\bheading{I Endpoint Identification} 
First, we identify entry point methods for our analysis. 
Thus, for web applications, we identify the request handling methods, i.e., endpoints, by using regular expressions to find functions with a \texttt{HttpRequest} parameter.

\bheading{II Global Slicing} 
Given the previously identified endpoints, we run forward slicing on a function granularity basis.
Initially, context-insensitive Control-Flow Analysis (0-CFA)~\cite{DBLP:conf/pldi/Shivers88/CFA} is used to quickly reduce the analysis scope. However, since this reduction may be too narrow and miss critical cases, we then use Class-Hierarchy Analysis (CHA)~\cite{DBLP:conf/ecoop/DeanGC95/CHA} to over-approximate and ensure correctness by covering all possible inheritance scenarios. Given all functions reachable per endpoint, we select those functions that contain a sink statement $s_l \in \mathcal{S}_{sink}$.
To determine which statements are potentially vulnerable, we rely on a well-established list~\cite{findsecbugs-injection-sink-list} by FindSecBugs~\cite{find-sec-bugs}.
For each function containing a sink, we create a backward slice. By taking the intersection of both slices, we retain only the functions on the path between an endpoint and a sink.

\bheading{III Tracing} 
Given the previously identified pairs of endpoints and sinks along with their functions in between, i.e., the scope, we statically generate all reachable traces. We utilize the JVM emulation capabilities of the symbolic backend provided by SWAT \cite{SWAT}, a dynamic symbolic execution engine for JVM bytecode. 
For our implementation, we developed a custom emulation engine that interacts with SWAT's execution tracking capabilities to enable accurate handling of both control (our engine) and data flow (SWAT). This also holds for some dynamic features (i.e., lambda expressions or dynamic invocations) for which we provide wrappers.  
Pairing the JVM emulation with a static instruction execution module allows us to statically emulate methods without having access to any specific parameters or values determined at runtime. 
During the static emulation, each conditional branch with an \textit{else} results in an independent duplication of the emulation. %
However, to ensure completeness and correctness we keep the branching statement and its condition in the \TG{}.
The static execution module records all observed instructions for each trace to assemble a single compiled method containing all statements observed during tracing.
While this creates an exponential number of traces in the branching depth, it does so only within the narrow limits imposed by the Global Slicing. %

\bheading{IV Local Slicing} 
To generate our Trace Gadgets from the previously obtained traces, we utilize a self developed static slicing engine.
Given the bytecode file of a trace, the engine slices the trace with respect to the potentially vulnerable sink. %

\bheading{V Optimization}   
To further optimize \TGs{}, we utilize the Proguard Optimizer~\cite{Proguard}.
This further reduces the number of statements. As we perform all operations on bytecode level, we generate Java source code as model input with Intellij's Fernflower decompiler~\cite{fernflower}. %

An example \TG{} generated from an OWASP Benchmark sample 
is shown in \Cref{lst:slice-example}. Here, code from different classes and functions are merged into a single function, and unnecessary computations are stripped, leaving a single method that encapsulates the essential functionality relevant to the sink.

In summary, our toolchain combines static tracing and slicing of arbitrary Jar files with automatic matching of potentially vulnerable statements as slicing criteria. Additionally, we utilize parametric endpoint identification to select an optimal entry point for tracing and perform an optimization pass to generate \textit{Trace Gadgets} for each endpoint and each possible execution path.
A detailed overview of the technical challenges and engineering contributions underlying Trace Gadget generation is provided in \Cref{appendix:engineering-challenges}.

\subsection{Limitations of the implementation}
\label{appendix:limitations_of_the_implementation}

Currently, we support all JVM bytecode instructions, except the \texttt{throw} instruction. 

\bheading{Exception Handling} The engine executes try-catch blocks without considering potential exceptions, therefore effectively ignoring catch blocks. Moreover, we can not handle execution paths that contain the \texttt{throw} instruction.
Upon experimental evaluation on real production programs (VulnDocker), this limitation is responsible for 4.5\% of endpoints not being emulated, as detailed in~\Cref{tab:eval:does-tracing-scale}.

\bheading{Internal Functionality} Currently, we only have limited wrapper functions for internal functionalities such as threading or reflection. The tool directly copies most of these functionalities without emulating their behavior. This approach succeeds for functionalities that do not alter program execution, such as string operations, but fails for, e.g., concurrency, as it does not replicate the execution order or resolve dynamic behaviors. 
A detailed description of the consequences of these limitations is given in \Cref{app:incorrect-semantics}.

%% file: sections/30-Dataset.tex
\section{Datasets}
\label{sec:datasets}

A diverse and high-quality dataset is essential for effective ML. If the training data misrepresents the target distribution or omits critical cases, the performance of applied algorithms suffers accordingly. For classifiers used beyond the training domain, the dataset must closely approximate real-world data, also implying the existence of both vulnerable and benign samples. %
Therefore, we created a new large-scale labeled dataset, \textit{VulnDocker}, based on applications extracted from docker containers hosted on DockerHub~\cite{dockerhub}. %
To ensure transferability, we decouple the datasets used for training (VulnDocker), hyperparameter evaluation (Juliet) and evaluation (OWASP). Evaluating our models on OWASP also ensures comparability to other vulnerability detection frameworks, as this dataset is commonly used for evaluation~\cite{DBLP:conf/compsac/Partenza, CodeQL, find-sec-bugs, slscan, Semgrep}.

\subsection{Benchmark Datasets}
\label{subsec:benchmark-dataset}

As described before, we use the Juliet and OWASP datasets, for hyperparameter evaluation (Juliet) and the overall evaluation (OWASP). 
Therefore, we preprocess the datasets and filter non-injection vulnerability related samples. 

For Juliet, many samples only differ marginally. Hence, after generating \TGs{} we substantially reduced redundancy in the dataset. In particular, we found that most were duplicates, resulting in only 587 unique \TGs{} from the original set of 8803 test cases. The final processed dataset contains $257$ benign and $330$ vulnerable \TGs{}.

For OWASP, due to the complexity of the dataset, most test cases resulted in multiple \TGs{} per endpoint. We obtain 5823 \TGs{} from a total of 1572 endpoints. 819 of those endpoints are vulnerable. 
As this dataset is used for evaluation only, a label per endpoint, rather than per \TG{}, is sufficient, as the ground truth of the OWASP dataset is endpoint-based. 
For evaluation purposes, we aggregate \TG{} classifications: an endpoint is deemed vulnerable if the maximum prediction score among all its associated \TGs{} surpasses a defined threshold. Only if all \TGs{} for an endpoint fall below this threshold (i.e., are classified as benign) the endpoint itself is classified as benign. This strategy enables direct performance assessment against the established endpoint-level ground truth.

\subsection{VulnDocker Dataset}
\label{subsec:our-dataset}
During the training phase, it is essential to utilize a robust dataset designed for ML applications. 
To allow for arbitrary analysis of JVM programs, including closed-source security analysis, the input to our framework is JVM bytecode. 
To obtain program code used in production environments, we explored DockerHub~\cite{dockerhub} as a source of real-world JVM applications. 
While GitHub offers a wide variety of projects, they require compilation with the correct production configuration. 
Moreover, compiling arbitrary GitHub projects is a difficult challenge on its own~\cite{50K-C}.
We collected an extensive list of $7\,459\,528$ unique containers and their respective description file. 
These description files provide a structured representation of the container's construction. 
We filter each based on its description for JVM applications, resulting in $9\,807\,555$ Jar files. 
Passing these files through the Trace Generation toolchain results in $182\,346$ unique traces and $32\,886$ unique \TGs{}. 
The main reason that led from such a large number of Jar files to a comparatively 
small number of \TGs{} was the presence of numerous duplicates. %

To train ML models, we needed labels for the collected \TGs{}. Since labeling code as vulnerable or benign is challenging without ground truth, we approximated labels using 
Find Security Bugs (FindSecBugs)~\cite{find-sec-bugs}, a state-of-the-art static scanner~\cite{DBLP:conf/xpu/OyetoyanMGC18/FindSecBugsIsTheBest}. %

However, since static scanners tend to overpredict~\cite{DBLP:conf/xpu/OyetoyanMGC18/FindSecBugsIsTheBest}, i.e., have a high false positive rate, we manually reviewed all $10\,610$ supposedly vulnerable samples.
To ensure accuracy, we followed the recommendation by Dorner \emph{et al.}~\cite{DBLP:conf/icml/DornerH24} and allocated effort to additional single reviews.
After this extensive process, we ended up with $1\,412$ truly vulnerable samples. 
Relying on FindSecBugs labels for benign samples introduces the risk of false negatives, where vulnerable code is mislabeled as benign. Therefore, we additionally reviewed a subset of 500 random benign samples. Of these, only 5 samples were identified as false negatives, corresponding to a false negative rate of 1\%. Thus, while we acknowledge the possibility that some vulnerable samples may be mislabeled as benign, such occurrences appear to be rare.

\par

%% file: sections/40-Evaluation.tex
\section{Experiments and Results}
\label{sec:evaluation}

In the previous sections, we introduced Trace Gadgets (\TGs{}) and VulnDocker with the goal to improve the ability to automatically scan applications for software vulnerabilities. In the following, we will address the research questions from \Cref{sec:research-questions}, to evaluate the suitability of \TGs{} for injection  vulnerability detection.

\subsection{LLM Performance and Code Context Size}
\label{sec:eval:llm-context-length}

Modern Large Language Models (LLMs) claim to support context lengths in excess of 128,000 tokens. However, recent studies on question-answering and long code context tasks indicate a performance degradation of LLMs as input size increases~\cite{DBLP:LLMs_tend_to_only_use_begin_and_end_of_text, DBLP:journals/corr/abs-2503-04359/long-code-benchmark, DBLP:journals/corr/abs-2505-07897/long-code-benchmark}.  To the best of our knowledge, this issue has not been studied in the context of security-critical code analysis tasks.
Thus, we adapt the experiment design of Liu \textit{et al.}~\cite{DBLP:LLMs_tend_to_only_use_begin_and_end_of_text} to vulnerability detection in source code, and measure how detection accuracy evolves with increasing number of code tokens.

We evaluate a diverse set of LLMs, including six open-source models, i.e., CodeLlama~13B (13b parameters), LLaMA~3.1~405B (405b parameters), LLaMA~3.3~70B (70b parameters), Mixtral~8x7B (8 experts each 7b parameters), Phi-4 (14b parameters), and WizardCoder~15B (15b parameters), and two commercial models, i.e., GPT-4o and Claude~3.7~Sonnet. %
We exclude chain‑of‑thought (CoT) and Retrieval-Augmented Generation (RAG) variants because they are orthogonal to our research question.
\notefelixi{Cite models}

\bheading{Experiment Design}
To systematically evaluate how LLMs handle increasing code context lengths, we use the CRUXEval dataset~\cite{DBLP:conf/icml/GuRLSS024/CRUXEval} and extend it for security tasks. Therefore, we manually construct a small dataset of vulnerable and secure functions.
Our dataset includes ten functions that exhibit common types of injection vulnerabilities, such as SQL injection, as well as their corresponding secure implementations. The LLM is tasked with identifying whether a given function contains a security vulnerability. Similar to Liu \emph{et al.}, we incrementally expand the prompt by inserting semantically independent distractor functions from the CRUXEval dataset, deliberately creating a conservative benchmark scenario with no inter-function dependencies. On each trial, the target function appears at a random position among the distractors, followed by the security analysis prompt. Function names were randomized across iterations, following the methodology of Liu \emph{et al.}, to prevent models from exploiting naming patterns. For every target function, both the vulnerable and the secure variant are included. Each model is queried five times per function, resulting in a total of 1,600 query calls per model.
Crucially, all evaluated models received identical target and distractor functions across corresponding experimental iterations, ensuring comparability.
Token counts were standardized across models using the GPT-4o tokenizer, ensuring consistent measurement despite internal tokenizer differences.

\begin{figure}[t]
    \centering
    \includegraphics[width=0.98\linewidth]{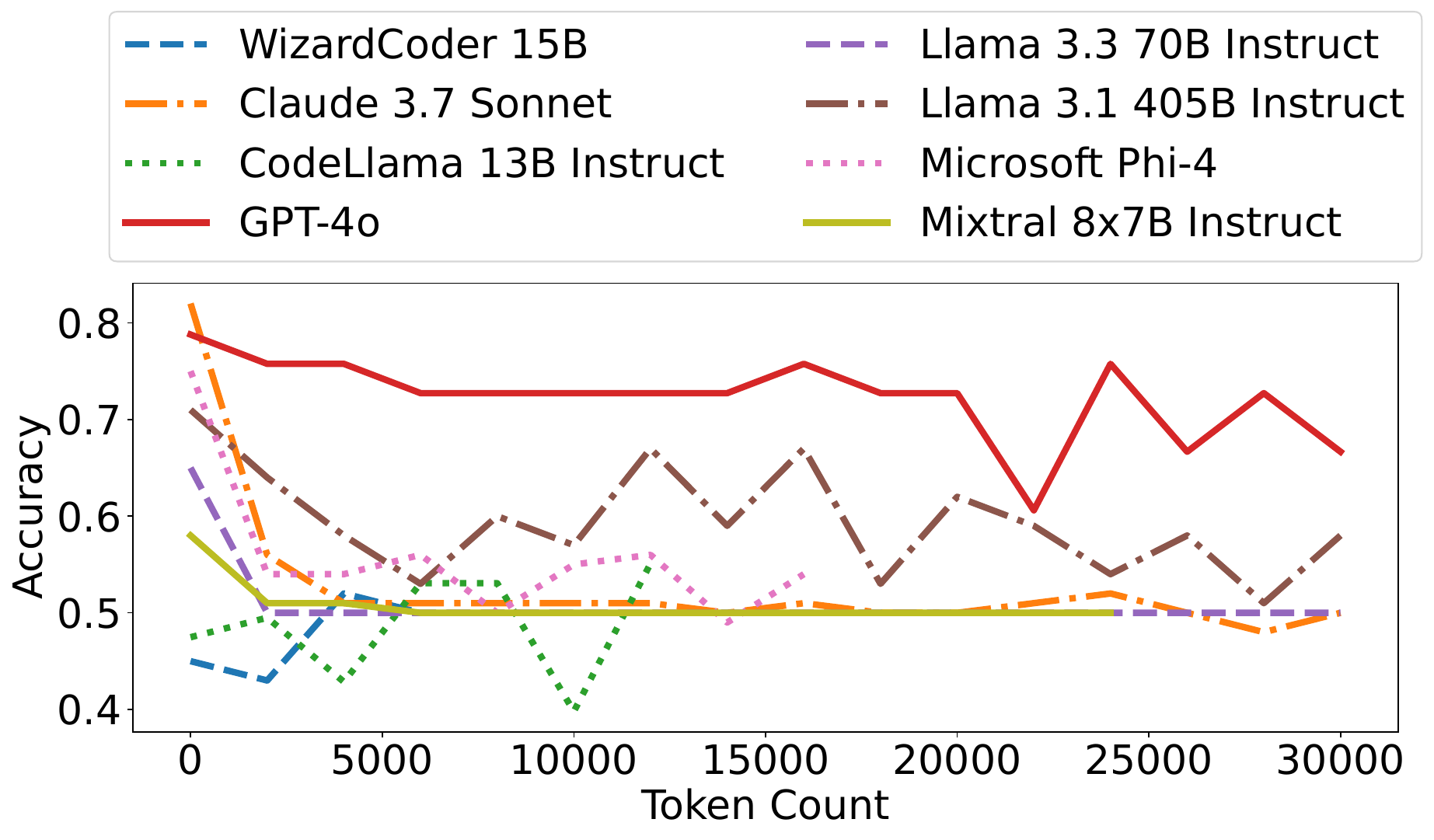}
    \caption{Success rate of various LLMs at predicting whether a function is vulnerable, depending on the context length of the input.\notefloi{I am really not happy about the legend hiding part of the plot.}\notefloi{Plot could be bigger.}}
    \label{fig:llm-context-length}
\end{figure}

\bheading{Results}
The detection accuracy of all evaluated LLMs on the vulnerability detection task against the token count of the input is shown in \Cref{fig:llm-context-length}.

At minimal context length, Claude~3.7~Sonnet achieves the highest baseline accuracy of 82.0\%, followed by GPT-4o (78.8\%), and Phi-4 (75.0\%).
As context length increases up to 30,000 tokens, all models exhibit performance degradation, albeit to varying degrees.
GPT-4o demonstrates the greatest robustness with only a 15.4\% relative decrease from 78.8\% to 66.7\%, maintaining functional security analysis capabilities even at extended context lengths.
In contrast, Claude~3.7~Sonnet, despite its superior baseline performance, experiences the most severe degradation with a 39.0\% relative decrease from 82.0\% to 50.0\%, effectively reducing accuracy to random guessing levels at maximum context. 
Among open-source models, Phi-4 exhibits a 28.0\% decrease, demonstrating moderate resilience despite being the strongest performer at baseline.
The remaining models %
show intermediate degradation patterns, though their generally modest baseline accuracies suggest these measurements may be predominantly influenced by noise rather than systematic context-length effects.

\bheading{Discussion}
These observations show that long contexts affect the accuracy of vulnerability detection across all models sizes.
Although modern LLMs theoretically support extensive context lengths, their performance on security-critical tasks drops when dealing with excessively long contexts. Our findings align with related work examining code-related reasoning under long-context conditions~\cite{DBLP:journals/corr/abs-2503-04359/long-code-benchmark, DBLP:journals/corr/abs-2505-07897/long-code-benchmark}.
This performance degradation is particularly concerning for security applications, as it suggests that LLMs may miss vulnerabilities when analyzing large codebases or when relevant code is embedded within extensive context.
The performance degradation is likely due to difficulties in maintaining relevant long-range dependencies, especially in the attention-layers~\cite{DBLP:journals/corr/abs-2410-01104}. 
Consequently, these findings underscore the importance of developing methods to minimize input size to maintain high model accuracy.

\begin{answerbox}
\textbf{\ref{item:rq0}:} Increasing context length has a negative impact on the vulnerability detection capabilities of LLMs.
\end{answerbox}

\subsection{Trace Gadget Generation Efficiency}
\label{sec:eval:tracegadetgeneration}
To answer \ref{item:rq1}, we first evaluate the efficiency of our Trace Gadget generation in this section as well as the overall quality of the \TGs{} in the next.
 \label{sec:eval:engine-performance}

\bheading{Experiment Design} To evaluate the efficiency of the trace generation, we consider two metrics: the proportion of endpoints for which \TGs{} can be successfully generated and the average and median time required for trace generation per endpoint. 
We evaluate these metrics on three datasets: OWASP, Juliet and VulnDocker.
For the Juliet and OWASP datasets, we consider all available endpoints. 
Due to the size of the VulnDocker dataset, we randomly sample $1000$ Jar files from the dataset and generate \TGs{} for one random endpoint of each Jar file.
The aggregated results for the success rate of Trace Gadget generation are shown in ~\Cref{tab:eval:does-tracing-scale}. 
\TGs{} could not be generated successfully if the processing time exceeds a timeout of $5$ minutes or if an error occurs within the engine. 

\bheading{Results} The performance across both benchmark datasets is excellent, with \TGs{} being successful generated for around $99\%$ of all endpoints.
However, for the VulnDocker dataset the success rate drops to 89.4\%. 
The 11.6\% of endpoints for which no \TGs{} could be generated can be attributed to 5.4\% timeouts and 5.2\% premature terminations. 
The 5.2\% of premature terminations are attributable to technical constraints, primarily due to shortcomings in our engine's handling of the JVM's \texttt{throw} instruction (4.5\%). 
The remaining 0.7\% of errors in the engine are caused by other limitations. %

To assess the time complexity, we analyzed the generation times of VulnDocker endpoints that produced at least one \TG{}.
We observed that on our machine equipped with a Neoverse-N1 CPU, the median time to generate a TG per endpoint is approximately $11.03\,s$ whereas the mean is $63.07\,s$. 

\input{figures/Training-Overview}

\input{tables/does-tracing-scale}
\bheading{Discussion} For TG generation on our real-world VulnDocker dataset we observe only 5.4\% of endpoints resulting in timeouts. Thus, the computational load of generating \TGs{} remains controlled. %
This control is primarily achieved by applying global slicing (Step II of our toolchain) prior to trace generation, which limits growth to paths from a single source to a single sink. Without global slicing, i.e., when all target functions are in scope, the timeout rate increases from 5.4\% to 45\% of all traces.

To assess the overall computation cost of \TG{} generation for a typical web application, we further evaluated the average number of endpoints per and the average number of \TGs{} per endpoint.
In the VulnDocker dataset, the number of endpoints per web application, are 4.9 in median and 57.54 on average. 
These endpoints result in a median of 4.0 and a mean of 24.41 \TGs{}. 
Therefore, combining the numbers with the median or average generation time of a TG, our analysis for a typical (median) web application is rather fast. 
Some applications take longer to generate \TGs{}, as indicated by the averages. However, even applications with twice the average number of endpoints complete in about an hour, which is significantly faster than a manual review by an expert.

\par

\subsection{Correctness of Trace Gadget Generation}
\label{rq:answer:Correctness of Trace Gadget Generation}
Evaluating the correctness of the generated \TGs{} is crucial for assessing the reliability and validity. %

\bheading{Experiment Design}
To assess the correctness of TG generation, we use $1000$ random test cases from a subset of the Juliet dataset.
The subset consists of all test cases %
 meeting the following criteria:
 \begin{itemize}[leftmargin=10pt, itemsep=0pt]
     \item When the test case is executed multiple times with a fixed random seed, the same output has to be observed.\notefloi{As in the original program? M: No, here we are selecting testcases from the original program. Hence, we execute the original program}
     \item The test case has to terminate within three seconds to safeguard against issues like infinite loops or hardware-damaging test cases, such as those designed to exploit 
     \textit{CWE-400 Resource Exhaustion}. 
     \item No memory addresses are allowed in the output. While the address is consistent across multiple executions, it is not identical to the addresses in \TGs{}.
 \end{itemize}

For each of these $1000$ test cases, we compare the output of the original test case with the output of the corresponding generated \TGs{}. If any TG's output matches the actual execution, we consider the semantics to be correct. 

\bheading{Results}
Our tracing engine successfully replicated the output in 937 instances, yielding a success rate of $93,7\%$. 
Upon manual inspection of the rest, we found that 38 cases were caused by known limitations by the engine as explained in \Cref{appendix:limitations_of_the_implementation}. A detailed analysis of the erroneous test cases is given in \Cref{app:incorrect-semantics}.

\label{rq:rq1}
\begin{answerbox}
\textbf{\ref{item:rq1}:} \TGs{} for JVM-based web applications can be efficiently generated with our framework. All context required for proper functionality is maintained.
\end{answerbox}

\subsection{Machine learning-based Vulnerability Detection}
\label{subsec:results-ml}
To answer \ref{item:rq2}, we first select three suitable models that report state-of-the-art performance on code comprehension tasks. 
After model selection, we present our experiment design for the training phase and report the performance of each model on our dataset. 

\bheading{Model Selection}
To evaluate the effectiveness of \TGs{}, we fine-tune a pre-trained ML model from literature using our prepared datasets. 
While related work~\cite{VulDeePecker, DeepWukong, sentence-encodings} presents freshly trained models or slightly improved architectures, we argue that the level of code-understanding achieved by pre-trained models cannot easily be achieved by training a model from scratch for the task at hand~\cite{Peculiar}. 
Consequently, we consider the design of a new architecture for the task of vulnerability detection out of scope and focus on fine-tuning models pre-trained with code understanding.
Especially, given the very precise but also very small dataset crafted for vulnerability detection, it is required to utilize a model that has been trained for general code understanding~\cite{Peculiar}. 
Although, all selected models have undergone pre-training on a range of tasks related to code comprehension, it is noteworthy that none of these tasks addressed vulnerability detection.

Specifically, we selected three state-of-the-art models for our evaluation.
We evaluate the performance of all three to determine the model best suited for our task.
All chosen models are transformer-based \cite{first_transformer}, and have been trained on a number of code-related tasks in order to gain code understanding.
Concretely, we picked \textit{UniXcoder}~\cite{DBLP:ACL2022_Guo_Unixcoder}, \textit{CodeT5+}~\cite{DBLP:Wang_CodeT5+} and \textit{Traced}~\cite{ICSE2024_traced_ding}, due to their  performance results~\cite{DBLP:Wang_CodeT5+, ICSE2024_traced_ding, OCEAN}.

\bheading{Experiment Design} 
Our training procedure consists of three training steps shown in~\Cref{fig:training-procedure}. %
This experiment focuses on the SQL-injection subsets of the evaluation datasets (i.e., OWASP and Juliet), as SQL injection attacks are the predominant vulnerability in the OWASP dataset. Moreover, Partenza \emph{et al.}~\cite{DBLP:conf/compsac/Partenza} also use the OWASP SQL injection subset.
We begin by fine-tuning a pre-trained model with our VulnDocker dataset, enabling it to understand the underlying code structures and common patterns of injection attacks.
To allow for hyperparameter selection we use a grid search with respect to the F1 score. We evaluate each set of hyperparameters on the subset of SQL injection examples of the unseen Juliet dataset. 
Applying an unseen dataset for hyperparameter selection enhances transferability.
More information regarding the selection of hyperparameters can be found in \Cref{app:hyperparams}.
Finally, we evaluate the fine-tuned models on the OWASP benchmark dataset due to its comparability with industry-grade scanners. 
To evaluate the performance on the OWASP dataset, we use the ground truth labels %
for each endpoint. 
If any TG within an endpoint is identified as vulnerable, we classify the entire endpoint as vulnerable. 
This approach allows us to evaluate the models' detection performance against the OWASP benchmark's endpoint-based vulnerability labels.

\bheading{Results}
\input{figures/Main-ML-Plot}
The previously conceptualized experiments were run against all three models. The F1 score of each model, including the standard deviation, is calculated over ten runs. 
All models were evaluated on an unused portion of the VulnDocker dataset, the SQL injection subset of the Juliet dataset, and the SQL injection subset of the OWASP dataset. The results are shown in~\Cref{fig:eval:main-ml-plot}. 
As expected, all models achieve the best performance on the training dataset with an average F1 score between $0.77$ and $0.81$.
The F1 scores reported for Juliet are expectantly a bit lower ranging from $0.69$ to $0.74$. 
When evaluating all models on the unseen OWASP dataset, the performance of Traced and CodeT5+ remains at a comparable level, with an average F1 score of $0.71$ for Traced and $0.73$ for CodeT5+. 
This highlights the generalizability of these two models under the given training procedure. 
With an F1 score of $0.59$, the trained UniXcoder model shows a notably lower performance. 
As UniXcoder and Traced share the same architecture, this can potentially be explained with additional pre-training tasks in Traced~\cite{ICSE2024_traced_ding}.

\label{rq:rq2}
\begin{answerbox}
\textbf{\ref{item:rq2}:} All three models show robust performance on datasets other than VulnDocker, with two models matching their performance between the hyperparameter selection dataset and the fully unseen dataset. %
\end{answerbox}

\subsection{Comparison with State-of-the-Art}

\label{sec:eval:detection-performance-comparison}

To evaluate the effectiveness of Trace Gadgets compared to other input representations presented in previous work, we selected two widely used static vulnerability detection approaches that exemplify different input representations. Although there are alternative approaches for vulnerability detection, the two major input representations are code slices or function-level input. Thus, we select two representative approaches for each input representation.
The first approach is VulDeePecker~\cite{VulDeePecker}, which pioneered the application of deep learning for vulnerability detection. The second approach by Shestov \emph{et al.}~\cite{DBLP:journals/access/ShestovLMMZCMTTK25} uses modern LLMs for the same task. 
To compare not only with academic work, we also evaluate four industry-leading static scanners in our experiment, i.e., ShiftLeft Scan~(v2.1.1)~\cite{slscan}, CodeQL~(v2.19.0)~\cite{CodeQL},  FindSecBugs~(v1.12.0)~\cite{find-sec-bugs} and Semgrep~(v1.86.0)~\cite{Semgrep}.
Complementing the evaluation with our own setup, we include our best model from the previous research question, i.e., CodeT5+ together with  \TGs{}.

\bheading{Experiment Design} 
We evaluate each approach using both its original input representation and \TGs{} to compare their performance. Additionally, we contrast the median number of tokens for all representations.

The authors of VulDeePecker did not publish their model nor their data preprocessing implementation. %
To compare VulDeePecker to our vulnerability detection approach, %
we use an open-source implementation of VulDeePecker~\cite{johnb110-VDPython}. 
To replicate the results presented by Shestov \emph{et al.}~\cite{DBLP:journals/access/ShestovLMMZCMTTK25}, who use function-level granularity with a fine-tuned LLM (WizardCoder~\cite{DBLP:conf/iclr/LuoX0SGHT0LJ24/WizardCoder}), we refrained from fine-tuning due to computational resource constraints and the substantial complexity associated with fine-tuning LLMs. Instead, we used GPT-4o, which, as shown in \Cref{sec:eval:llm-context-length}, achieves significantly better performance compared to WizardCoder. Details about this setup can be found in \Cref{appendix:llm-prompt}. Additionally, in order to evaluate all code representations consistently, GPT-4o was tested with VulDeePecker's CodeGadgets.

Code Gadgets were extracted with WALA~\cite{wala} and functions with Tree-Sitter~\cite{tree-sitter}.
Consistent with our approach in the previous section, all approaches requiring training were trained on the Juliet dataset. All evaluations in this section were performed on the SQL injections part of the OWASP Benchmark. 
VulnDocker could not be used for training as it consists only of Trace Gadgets. Extracting other representations would have required rerunning the entire pipeline for a large amount of Jar files, including the manual verification of all labels. However, we argue that using the same dataset for training allows for a fair comparison between the input representations.

The results for all static scanners were retrieved using the scripts included with the OWASP benchmark. 
The metrics for all evaluations are the True Positive Rate (TPR) and False Positive Rate (FPR). 
These are the suggested metrics for the OWASP dataset~\cite{owasp-java}. We classify samples using a trivial decision threshold $\tau = 0.5$. An analysis of alternative thresholds is provided in \Cref{appendix:classification-thresholds}.
The results are shown in~\Cref{fig:eval:owasp-benchmark-sqli-only}. 

\bheading{Performance}
When the VulDeePecker model is trained with CodeGadgets, it achieves a moderate performance, i.e. a True Positive Rate (TPR) of 0.96 and a False Positive Rate (FPR) of 0.93.  
However, when CodeGadgets are replaced with \TGs{}, VulDeePecker's performance improves significantly. Although the TPR drops slightly to 0.84, the FPR drops dramatically to 0.68, a 29\% decrease. %

Comparing the results of the approach by Shestov \emph{et al.}, we find that combining Code Gadgets as well as function-level granularity as input representation together with GPT-4o, both result in a TPR and FPR of 1, which is as strong as random guessing.
However, replacing the representation with \TGs{} significantly improves performance, achieving a TPR of 0.95 and an FPR of 0.62, thereby reducing the FPR by 38\%.%

The static scanners ShiftLeft Scan, CodeQL, and FindSecBugs have excellent TPRs of $1.00$. However, this high sensitivity comes at the cost of elevated false positive rates: ShiftLeft Scan reports an FPR of $0.81$, CodeQL yields $0.89$, and FindSecBugs $0.90$.
For readability reasons, only the best-performing scanner, i.e., Semgrep, is shown in~\Cref{fig:eval:owasp-benchmark-sqli-only}. In comparison to ShiftLeft Scan, Semgrep achieves a slightly better result, 
with a TPR of $0.86$ and FPR of $0.60$.

Upon comparing the best static scanner, i.e., Semgrep, and the best ML based scanner, i.e., \TGs{} combined with GPT-4o, with our approach, i.e., CodeT5+ trained on VulnDocker, we see that Semgrep achieves the worst F1 score of 0.74, whereas  CodeT5+ with \TGs{} has a higher FPR of $0.73$, but also a substantially higher TPR of $1$, resulting in a F1 score of $0.76$. The combination of \TGs{} with GPT-4o outperforms all other scanners with an F1 score of $0.77$, beating Semgrep's F1 score by 4\% and slightly exceeding CodeT5+'s F1 score by 1\%.

\bheading{Token count} The use of \TGs{} not only improves detection accuracy but also results in the lowest token count after preprocessing. When comparing the preprocessed token count of the Juliet dataset, we observe that function-level granularity, despite lacking complete information, produces the highest number of tokens, with a median of 178. 
Code slices, such as CodeGadgets utilized by VulDeePecker, have a median token count of 165. 
On the other hand, \TGs{}, have the lowest median token count of 118. 
Thus, \TGs{} reduce the token count by 28--34\% compared to existing approaches.

\bheading{Discussion}
Although GPT-4o may have encountered the OWASP dataset during training, our results show that using unmodified function-level code, which is closer to GPT-4o's training data, results in worse detection performance compared to highly restructured \TGs{}. 
A plausible explanation is that the OWASP dataset contains many apparent vulnerabilities that are unreachable due to broken flows. Such broken flows may not be adequately detected by GPT-4o. On the other hand, \TGs{} implicitly capture those broken flows that render many vulnerabilities in the OWASP dataset inaccessible, thus substantially reducing the FPR.

Furthermore, it is notable that the superior performance of \TGs{} with GPT-4o is not caused by the commercial model, but by the code representation: 
Upon comparing the different input representations (A, B, E) in~\Cref{fig:eval:owasp-benchmark-sqli-only}, it can be seen that GPT-4o with functions and slices results in a worse detection performance than with \TGs{}.

\input{figures/OWASP-TPR-FPR}
\begin{answerbox}
\textbf{\ref{item:rq5}:} \TGs{} improve the detection performance of ML-based vulnerability detection, compared to Code Gadgets or function-level granularity. %
Moreover, \TGs{} contain the fewest statements while being complete.
\end{answerbox}

\subsection{Trace Gadgets and the Code Representation Requirements}
\label{sec:rq:answer:do_trace_gadgets_fulfill_requirements}
In~\Cref{sec:code-representation} we identified five requirements for code representations that have to be met for vulnerability detection.
To assess whether \TGs{} fulfill all five requirements, we revisit each and check whether they are fulfilled:

\bheading{\reqOne} By including only statements necessary for a single execution path, \TGs{} are highly concise, thus satisfying the conciseness requirement. This is underlined by the results in~\Cref{sec:eval:detection-performance-comparison}, where \TGs{} contain the fewest statements compared to other code representations.

\bheading{\reqTwo} Despite their conciseness, \TGs{} are complete because they contain all necessary statements, as shown in~\Cref{rq:answer:Correctness of Trace Gadget Generation}, by successfully replicating most program outputs. In particular, each execution path is captured by a single \TG{}, so the union of all \TGs{} covers the entire program.

\bheading{\reqThree} By design, \TGs{} only consist of a single function representing the execution trace. Hence, they also fulfill the simplicity requirement.

\bheading{\reqFour} The use of an optimizer and a decompiler in the generation of \TGs{} standardizes the naming and coding style. This uniformity eliminates stylistic variations, satisfying the requirement for stylistic consistency.

\bheading{\reqFive} While the approach theoretically %
allows for exponential growth, in practice the \TG{} generation process remains computationally efficient, with a median generation time of $11\,s$ per trace, as demonstrated in \Cref{sec:eval:tracegadetgeneration}.

\begin{answerbox}
    \textbf{\ref{item:rq4}} {\TraceGadget}s meet all %
    requirements.
\end{answerbox}

\subsection{Evaluation on Real-World Targets}
\label{subsec:vuln-findings}
To evaluate our proposed setup beyond laboratory conditions, we applied it in real-world scenarios. %

\bheading{Experiment Design} 
From the list of Docker containers as described in \Cref{subsec:our-dataset} we select those containers with over one million pulls and extract their Jar files. 
For those Jar files, we applied our described pipeline from \Cref{fig:system-overview} together with a CodeT5+ model trained on VulnDocker.

\bheading{Results}
Our investigation revealed two previously unknown vulnerabilities in popular Java-based web services. %

First, we identified a cross-site scripting (XSS) vulnerability in Atlassian Bamboo, a widely deployed build server, with over 1.5 million pulls on DockerHub. 
Specifically we discovered a stored XSS vulnerability resulting in a privilege elevation. 
The vulnerability allows the creation of a new administrator account.
After we reported this vulnerability, it was acknowledged and fixed~\cite{Bugcrowd-Finding1} by the vendors.
Our second discovery was a server-side request forgery (SSRF) in Geoserver, an open-source project for sharing and editing geospatial data.
This vulnerability has been concurrently reported by us and an independent security researcher. The vendor acknowledged both reports and released a fix~\cite{patch-geoserver}.

%% file: figures/Training-Overview.tex
\begin{figure*}[t]
    \centering
\includegraphics[scale=.75]{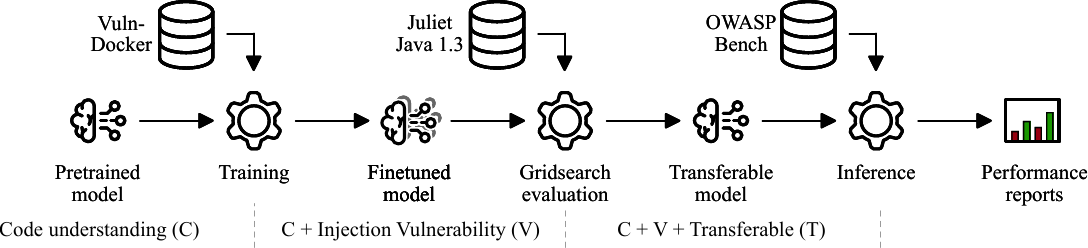}
    \caption{Systematic overview of the training procedure (middle) alongside all utilized datasets (top) and the model capabilities (bottom).} %
    \label{fig:training-procedure}
\end{figure*}

%% file: tables/does-tracing-scale.tex
\begin{table}[t]
    \centering
    
    \caption{
    Trace Gadget generation results in proportion to the number of endpoints. The ``Throw'' and  ``Error'' columns refer to limitations of the engine.%
    }
    \label{tab:eval:does-tracing-scale}
    \begin{threeparttable}
        \begin{tabularx}{\linewidth}{Xcccc}
        \toprule
        & Successful &  \multicolumn{3}{c}{Unsuccessful} \\
        \midrule
        Dataset    &  & Timeout & Throw & Error \\ 
        \midrule

        VulnDocker & 0.894 & 0.054 & 0.045 & 0.007 \\
        Juliet & 0.989 & 0.011 & 0 & 0  \\
        OWASP & 0.991 & 0.009 & 0 & 0 \\
        
        \bottomrule
        \end{tabularx}
    \end{threeparttable}
    
\end{table}

%% file: figures/Main-ML-Plot.tex
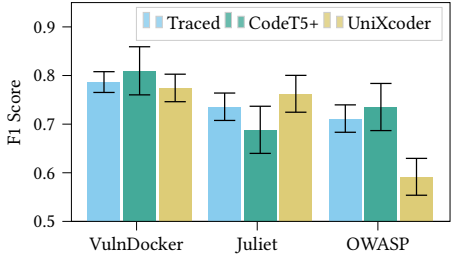
\begin{figure}[t]
    \centering
\input{assets/Tikz/ml-compare-models-plot}$\hspace{1em}$
    \caption{The F1 scores for each model across all three datasets across 10 runs. VulnDocker was used as  training dataset and the hyperparameters of the models were tuned using the Juliet dataset. Consequently, the OWASP dataset represents an entirely unseen dataset in this evaluation. %
    }
    \label{fig:eval:main-ml-plot}
\end{figure}

%% file: assets/Tikz/ml-compare-models-plot.tex
\begin{tikzpicture}[scale=.8]

\definecolor{burlywood221204119}{RGB}{221,204,119}
\definecolor{cadetblue68170153}{RGB}{68,170,153}
\definecolor{darkgray176}{RGB}{176,176,176}
\definecolor{lightgray204}{RGB}{204,204,204}
\definecolor{skyblue136204238}{RGB}{136,204,238}

\begin{axis}[
width=\linewidth-2em,
height=5.2cm, %
legend cell align={left},
legend columns=3,
legend style={fill opacity=0.8, draw opacity=1, text opacity=1, draw=lightgray204},
tick align=outside,
tick pos=left,
x grid style={darkgray176},
xmin=-0.27, xmax=2.868,
xtick style={color=black},
xtick={0.27,1.27,2.27},
xticklabels={VulnDocker,Juliet,OWASP},
y grid style={darkgray176},
ylabel={F1 Score},
ymin=0.5, ymax=0.95,
ytick style={color=black},
ytick={0.5,0.6,0.7,0.8,0.9},
yticklabels={0.5,0.6,0.7,0.8,0.9}
]
\draw[draw=none,fill=skyblue136204238] (axis cs:-0.135,0) rectangle (axis cs:0.135,0.786531471590169);
\addlegendimage{ybar,ybar legend,draw=none,fill=skyblue136204238}
\addlegendentry{Traced}

\draw[draw=none,fill=skyblue136204238] (axis cs:0.865,0) rectangle (axis cs:1.135,0.735769597088563);
\draw[draw=none,fill=skyblue136204238] (axis cs:1.865,0) rectangle (axis cs:2.135,0.711462751876561);
\draw[draw=none,fill=cadetblue68170153] (axis cs:0.16,0) rectangle (axis cs:0.43,0.809685752515014);
\addlegendimage{ybar,ybar legend,draw=none,fill=cadetblue68170153}
\addlegendentry{CodeT5+}

\draw[draw=none,fill=cadetblue68170153] (axis cs:1.16,0) rectangle (axis cs:1.43,0.68832253715916);
\draw[draw=none,fill=cadetblue68170153] (axis cs:2.16,0) rectangle (axis cs:2.43,0.735215439664997);
\draw[draw=none,fill=burlywood221204119] (axis cs:0.455,0) rectangle (axis cs:0.725,0.774444451242028);
\addlegendimage{ybar,ybar legend,draw=none,fill=burlywood221204119}
\addlegendentry{UniXcoder}

\draw[draw=none,fill=burlywood221204119] (axis cs:1.455,0) rectangle (axis cs:1.725,0.762417584460187);
\draw[draw=none,fill=burlywood221204119] (axis cs:2.455,0) rectangle (axis cs:2.725,0.591681015487967);
\path [draw=black, semithick]
(axis cs:0,0.765163013365549)
--(axis cs:0,0.80789992981479);

\path [draw=black, semithick]
(axis cs:1,0.707596667272132)
--(axis cs:1,0.763942526904994);

\path [draw=black, semithick]
(axis cs:2,0.68328982206013)
--(axis cs:2,0.739635681692992);

\addplot [semithick, black, mark=-, mark size=5, mark options={solid}, only marks]
table {%
0 0.765163013365549
1 0.707596667272132
2 0.68328982206013
};

\addplot [semithick, black, mark=-, mark size=5, mark options={solid}, only marks]
table {%
0 0.80789992981479
1 0.763942526904994
2 0.739635681692992
};

\path [draw=black, semithick]
(axis cs:0.295,0.760214594134298)
--(axis cs:0.295,0.859156910895729);

\path [draw=black, semithick]
(axis cs:1.295,0.639854964260514)
--(axis cs:1.295,0.736790110057806);

\path [draw=black, semithick]
(axis cs:2.295,0.686747866766351)
--(axis cs:2.295,0.783683012563643);

\addplot [semithick, black, mark=-, mark size=5, mark options={solid}, only marks]
table {%
0.295 0.760214594134298
1.295 0.639854964260514
2.295 0.686747866766351
};

\addplot [semithick, black, mark=-, mark size=5, mark options={solid}, only marks]
table {%
0.295 0.859156910895729
1.295 0.736790110057806
2.295 0.783683012563643
};

\path [draw=black, semithick]
(axis cs:0.59,0.74616204983164)
--(axis cs:0.59,0.802726852652417);

\path [draw=black, semithick]
(axis cs:1.59,0.724520881013273)
--(axis cs:1.59,0.800314287907101);

\path [draw=black, semithick]
(axis cs:2.59,0.553784312041053)
--(axis cs:2.59,0.62957771893488);

\addplot [semithick, black, mark=-, mark size=5, mark options={solid}, only marks]
table {%
0.59 0.74616204983164
1.59 0.724520881013273
2.59 0.553784312041053
};

\addplot [semithick, black, mark=-, mark size=5, mark options={solid}, only marks]
table {%
0.59 0.802726852652417
1.59 0.800314287907101
2.59 0.62957771893488
};

\draw [black, thin] (axis cs:-0.27,0.5) -- (axis cs:2.868,0.5);

\end{axis}

\end{tikzpicture}

%% file: figures/OWASP-TPR-FPR.tex
\begin{figure}[t]
    \centering

    \includegraphics[width=0.8\linewidth]{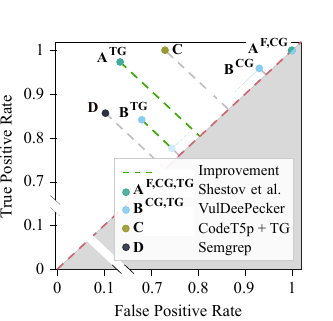}
    
    $\vspace{-2em}$\caption{%
    True Positive Rate (TPR) and False Positive Rate (FPR) on the OWASP benchmark with clipped axis. %
    The letters A, B, D refer to the approaches from related work and C to our approach from RQ3. Superscripts indicate the code representation, i.e. functions (F), Code Gadgets (CG), or our Trace Gadgets (TG). Green dashed lines show how switching to \TGs{} improves performance.%
    }
    
    \label{fig:eval:owasp-benchmark-sqli-only}
\end{figure}

%% file: sections/50-related-work.tex
\section{Related Work}
\label{sec:realted-work}

Vulnerability detection has a long history in the security community. Various code analysis methods from different areas have been applied to this challenge, ranging from rule-based approaches such as taint analysis~\cite{DBLP:conf/pldi/ArztRFBBKTOM14/FlowDroid}, to formal methods such as symbolic execution~\cite{DBLP:journals/cacm/King76}, random testing through fuzzing~\cite{fuzzingbook2024}, and function similarity~\cite{DBLP:conf/sp/YamaguchiGAR14}.
In the last decade, most of these methods have seen success through their joint application with machine learning techniques~\cite{VulDeePecker, DBLP:conf/pldi/0022S20/LiGer/Traces, DBLP:conf/icse/WangTTWLFX024/Concoction, DBLP:conf/uss/MirskyMBYPDML23/VulChecker}. In particular, the use of artificial code understanding can bridge the gap to accurate vulnerability classification based on representations extracted by classical analysis methods.
As a result, two orthogonal areas of improvement have emerged: one focused on improving code representations through targeted refinements to code analysis methods, and the other focused on enhancing the capabilities of machine learning models by optimizing their architecture, size, and training procedures.

\bheading{Program Slicing} Li \emph{et al.} introduce deep learning-based vulnerability detection~\cite{VulDeePecker}. 
As an input representation, they use Code Gadgets, which are static program slices extracted with respect to potentially vulnerable API calls.
Code Gadgets capture the data-flow dependencies leading to a vulnerable statement, but include multiple possible execution paths and thus extraneous statements.%
Several works such as SySeVR~\cite{SySeVR} or Snopy~\cite{DBLP:conf/kbse/Cao000B00L024/VFC_slicing_back_and_for} extended upon Code Gadgets by computing both forward and backward slices. 
Although such extensions ensure that both causes and consequences of vulnerabilities are included, they may be redundant, especially for injection vulnerabilities where forward slices (beyond the sink) add little value because the injection has already occurred. 
Other approaches explored custom slicing strategies tailored to the vulnerability type, such as combining multiple sinks with forward and/or backward slicing~\cite{DBLP:conf/sigsoft/WuLXLS023}.
Unlike traditional slicing approaches that capture multiple paths, \TGs{} only contain a single execution path, %
thereby eliminating irrelevant code. 
In addition, \TGs{} produce a single, inlined function rather than multiple functions or classes. %

\bheading{Graph-based Models} Other works, e.g.,  DeepWukong~\cite{DeepWukong} or GLICE~\cite{DBLP:conf/eurosp/KrakerVH23/Glice}, use slicing  but change the underlying model architecture to graph-based models.
Additionally, there are numerous extensions to improve graph-based vulnerability detection, such as custom encodings of the graph's nodes~\cite{DBLP:conf/icse/SteenhoekGL24}, retrieval of graph nodes corresponding to vulnerable code lines~\cite{DBLP:conf/uss/MirskyMBYPDML23/VulChecker, DBLP:conf/sigsoft/Li0N21} or the combination of textual and graph representations~\cite{DBLP:journals/sensors/ZhangHC24/use_graph_and_code_as_input, Peculiar}.
Moreover, Wu \emph{et al.}~\cite{Peculiar} discovered that pretraining is beneficial for the task at hand. Hence, our decision to utilize pretrained models for our evaluation. 
The interested reader is refereed to ~\cite{DBLP:conf/internetware/ZhangZYWJ23/ComparingDifferentCodeRepresentations} for a comparison between different model architectures for the task of vulnerability prediction. 
The change of model architecture is, however, orthogonal to our approach. Trace Gadgets could serve as an intermediate preprocessing step for graph-based models, effectively reducing the graph size and aiding with \reqOne.

\bheading{Dynamic Execution Traces} %
The authors of LiGer~\cite{DBLP:conf/pldi/0022S20/LiGer/Traces} adopt a hybrid approach by using both symbolic execution and dynamic execution traces. 
By combining these two sources, the authors achieve state-of-the-art results in name prediction and semantic classification. 
Similar to LiGer, Concoction~\cite{DBLP:conf/icse/WangTTWLFX024/Concoction} uses dynamic symbolic traces in conjunction with function-level code. %
Both tools use their generated representations as input %
for ML-based classification.

In contrast, our method statically derives a single new function that represents the compressed execution, without noise and redundancy introduced by capturing all executed statements.
Moreover, \TGs{} could also complement Concoction by replacing function-level source code in their workflow with \TGs{}, thereby additionally  satisfying  \reqOne{} and \reqFour{}.%

\bheading{Function-Level Representations}
Many approaches~\cite{DBLP:conf/msr/FuT22/LineVul, DBLP:conf/icse/SteenhoekGL24, DBLP:conf/nips/ZhouLSD019/Devign, DBLP:Achilles, sentence-encodings, DBLP:conf/compsac/Partenza, DBLP:conf/nips/ZhouLSD019/Devign} rely on using entire functions as input representation.  However, as shown in \Cref{tab:code-representation:comparison-table}, function-level representation only satisfies two of the five requirements for code representations. 

\subsection{LLM-based Vulnerability Detection}

The rise of Large Language Models (LLMs) has opened new avenues for vulnerability detection. 
Recent works by Khare \emph{et al.}~\cite{DBLP:journals/corr/abs-2311-16169/OWASP-LLM} and Shestov \emph{et al.}~\cite{DBLP:journals/access/ShestovLMMZCMTTK25} use LLMs, either off-the-shelf or fine-tuned, to predict vulnerabilities. 
This was extended by Li \emph{et al.}~\cite{DBLP:conf/iclr/Li0N25}, who equip LLMs with tools to search the code base, and by Ding \emph{et al.}~\cite{DBLP:conf/nips/DingPMKYR24} and Wang \emph{et al.}~\cite{DBLP:conf/nips/WangZSXX024}, who prompt LLMs to reason about code execution semantics.
However, all these methods use conventional code inputs, such as full functions or slices. We, however, focus on a novel code representation independent of the ML model used.
Additionally, our experiments (\Cref{sec:eval:llm-context-length}) indicate that even modern LLMs perform better with shorter input sizes and are highly sensitive to the quality of the input representation. In contrast, \TGs{} provide concise, execution-aware, and semantically complete inputs, making them orthogonal to the above.

\subsection{Java Vulnerability Detection}

\bheading{With Data Snooping} When focusing on JVM-based languages, 
Achilles~\cite{DBLP:Achilles} is one of the first deep learning systems for Java. 
It predicts vulnerabilities with a function-level input %
with a Long Short-Term Memory network (LSTM)~\cite{DBLP:journals/neco/HochreiterS97/LSTM}. %
ISVSF~\cite{sentence-encodings} works similar to Achilles, but vectorizes each basic block instead of each token.

All previously described systems rely on a single dataset for training and evaluation, thus risking Data Sooping~\cite{DosAndDonts}. We depart from this approach, ensuring robust results and transferability. 

\bheading{Without Data Snooping} In contrast, Partenza \emph{et al.}~\cite{DBLP:conf/compsac/Partenza} did not train and evaluate on the same dataset. 
Their approach uses a 
sequential version of the program's Abstract Syntax Tree (AST) together with the corresponding source code.
After training on Juliet, they evaluated this model using the SQL injection subset of the OWASP Benchmark dataset. 
While having an excellent F1 score of $0.93$ in recognizing SQL injections in the Juliet dataset, they only gain a F1 score of $0.57$ on the SQL injection subset. 
Hence, our F1 score on the unseen OWASP Benchmark of $0.76$ / $0.77$ (\Cref{sec:eval:detection-performance-comparison}) reflects a performance improvement of $33 - 35\%$.
Similarly, Mamede \emph{et al.}~\cite{DBLP:conf/qrs/Mamede/2022} used a pre-trained version of CodeBERT~\cite{DBLP:conf/emnlp/FengGTDFGS0LJZ20/CodeBert} to identify vulnerabilities using the function's source code and categorize the vulnerabilities according to their Common Weakness Enumeration (CWE). 
While their approach showed impressive performance achieving F1 scores as high as 0.94 on the Juliet dataset, they reported a significant performance drop of 50-70\% in F1 scores when evaluated on a custom, real-world dataset. 
Unlike all the aforementioned approaches, our approach does not require the source code of the target because it works on bytecode. %

%% file: sections/60-Threats-to-Validity.tex
\section{Limitations}
\label{sec:threads-to-validity}

\noindent\bheading{Language Agnosticism} Our implementation and the resulting VulnDocker dataset are currently focused on JVM-based languages, as the prototype works directly on JVM bytecode. While languages such as C/C++ are not included, our approach is generally applicable to any programming language with source-to-sink vulnerabilities and analyzable control and data flow. Adapting the approach to other languages would primarily require a reimplementation of core analysis components, a task whose feasibility is supported by the availability of such analysis equivalent tools~\cite{joern}.

\noindent\bheading{Independent Distractor Functions}
Our long-context experiments in \Cref{sec:eval:llm-context-length}  use independent distractor functions to isolate the effect of context size, following the CRUXEval methodology. While effective for controlled measurement, this setup diverges from real software, where functions interact through shared state and call dependencies. Thus, the observed degradation likely underestimates the effect in realistic settings, where additional context may introduce stronger semantic interference.

\noindent\bheading{Generalization Beyond Injection Vulnerabilities}
Trace Gadgets currently target source--to--sink injection vulnerabilities, which naturally align with path-based data-flow analysis. Extending \TGs{} to vulnerability classes such as authorization or logic flaws is possible but nontrivial, as these often depend on predicate checks rather than explicit data flows. Capturing such patterns would require the model to learn control-dependence and check-use relationships, e.g., whenever a user database table is updated, a check to ensure that the user is authorized, must be performed first.

\noindent\bheading{Adversarial Manipulation and Data Poisoning}
Like other analysis-driven code representations, \TGs{} are susceptible to adversarial manipulation~\cite{DBLP:conf/dimva/LooseMPB023/Madvex}. An attacker could craft training samples that bias the learned distribution or construct adversarial code variants that evade detection at inference time. Although such threat models are beyond our present scope, future work should investigate hardening both model training and inference. %

%% file: sections/70-Appendix.tex
\section{Appendix}

\subsection{Evaluating Trace Gadgets Semantics}
\label{app:incorrect-semantics}
In~\Cref{rq:answer:Correctness of Trace Gadget Generation}, we discuss the observed discrepancies between the output of the generated Trace Gadgets and the original test cases. The following section provides a detailed analysis of all erroneous cases, highlighting two prevalent issues:

The primary problem, affecting 27 test cases, was the use of try-catch blocks in the original test cases. For example, the generated Trace Gadget crashed when attempting to load a configuration file that did not exist on our test system. In contrast, the original test case defaulted to a catch block. Our current engine does not encompass the handling of catch-blocks, leading to their absence in the Trace Gadgets generated. Upon manual inspection, it was confirmed that while the generated code was semantically accurate, it lacked the necessary catch-blocks.
Another 17 test cases produced different outputs because of issues in fixing the random crypto seed. As a result, while they correctly encrypted a string, the final ciphertext was different. However, we confirmed that despite the different output, the generation process was the same.
The remaining 19 test cases failed due to minor problems. Five were related to the loss of default initialization values in variables, resulting in incorrect output. Three cases involved threading, which our system currently fails to handle.
The missing 11 are attributed to the inability to emulate Javas internal behavior as discussed in~\Cref{appendix:limitations_of_the_implementation}. Given the rarity of this problem, we have decided not to modify our slicing engine at this time, although it is recognized as an area for future improvement.

\subsection{LLM Prompt}
\label{appendix:llm-prompt}
The experiments in \Cref{sec:eval:detection-performance-comparison} involved the use of LLMs, specifically OpenAI's GPT-4o, version gpt-4o-2024-05-13. For reproducibility purposes, the following prompt was used with a temperature of 0:

\begin{lstlisting}
system:
You are an expert in sql injection detection

human:
As an expert in vulnerability detection, your task is to analyze
the provided code snippet for the presence of CWE-89 
SQL Injection vulnerabilities.
            
Task:
- Focus exclusively on identifying CWE-89 SQL Injection issues.
- Do not address any other vulnerabilities or code issues.
- Provide a clear and concise assessment stating whether the 
  code contains CWE-89 SQL Injection vulnerabilities.
- If vulnerabilities are found, explain where they occur and why 
  they are vulnerabilities.
            
Code Snippet:
{code_snippet}
\end{lstlisting}

\subsection{Hyper‑parameter Grid and Search Protocol}
\label{app:hyperparams}

To ensure a fair comparison, we applied an identical grid–search procedure
to all models, i.e., \textsc{UniXcoder}, \textsc{Traced},
and \textsc{CodeT5+}.  The Cartesian product of the values below
generates \(3 \times 3 \times 3 \times 2 \times 2 = 108\) configurations
per model. Each model was trained on \textit{VulnDocker} and evaluated on the
\textit{Juliet} validation split. The setting with the highest
\(\mathrm{F_{1}}\) score was finally deployed on the OWASP benchmark with a higher number of seeds.

\begin{tabular}{@{}ll@{}}
    \toprule
    \textbf{Hyper‑parameter} & \textbf{Values}\\
    \midrule
    Learning rate           & \(\{1\times10^{-5},\,5\times10^{-5},\,1\times10^{-4}\}\)\\
    Drop‑out rate           & \(\{0.1,\,0.2,\,0.3\}\)\\
    Warm‑up fraction        & \(\{0,\,0.1,\,0.2\}\) (as proportion of total steps)\\
    Frozen encoder layers\(^\dagger\) & \(\{9,\,6\}\)\\
    Random seed             & \(\{23,\,42\}\)\\
    \bottomrule
\end{tabular}
\raggedright
  \(^\dagger\)Layer count starts at the embedding layer; a higher number
  freezes more lower layers during fine‑tuning.

\subsection{Impact of the Classification Threshold}
\label{appendix:classification-thresholds}

\begin{table}[h]
\centering
\caption{F1 scores of CodeT5+ on the SQL injection subset of the OWASP benchmark dataset for different classification thresholds $\tau$.}
\label{tab:threshold-sweep}
\begin{tabular}{lccccccccc}
\toprule
$\tau$ & 0.1 & 0.2 & 0.3 & 0.4 & 0.5 & 0.6 & 0.7 & 0.8 & 0.9 \\
\midrule
F$_1$            & 0.71 & 0.71 & 0.71 & 0.71 & 0.76 & 0.00 & 0.00 & 0.00 & 0.00 \\
\bottomrule
\end{tabular}
\end{table}

We evaluated the Trace Gadget--based classifier on the OWASP benchmark for thresholds $\tau \in \{0.1, 0.2, \dots, 0.9\}$, classifying a sample as vulnerable if its predicted probability exceeds $\tau$. The resulting F$_1$ scores are shown in Table~\ref{tab:threshold-sweep}: for $\tau \leq 0.4$ the F$_1$ score remains stable at 0.71, it peaks at 0.76 for the trivial threshold $\tau = 0.5$, and collapses to 0.0 for $\tau \geq 0.6$.

The collapse is a direct consequence of the definitions of precision, recall, and F$_1$: as $\tau$ increases, fewer samples are predicted as vulnerable, and in our setting almost all true vulnerable samples receive scores only slightly above 0.5. Once $\tau \geq 0.6$, the number of true positives (TP) effectively drops to zero, implying precision $= \frac{\text{TP}}{\text{TP} + \text{FP}} = 0$, recall $= \frac{\text{TP}}{\text{TP} + \text{FN}} = 0$, and therefore F$_1 = 0$. We attribute the concentration of positive scores near $0.5$ to a calibration shift between the training data (VulnDocker/Juliet) and the evaluation data (OWASP). Within this regime, $\tau = 0.5$ is empirically near-optimal. Notably, this threshold sweep was conducted only after all other experiments were completed, to verify whether the trivial value $\tau = 0.5$ is indeed optimal. Thus, no data snooping occurred.

\subsection{Engineering Challenges}
\label{appendix:engineering-challenges}

Implementing Trace Gadgets required substantial engineering efforts. Although \TGs{} conceptually build upon classical ideas, constructing a fully static JVM-level emulation framework capable of producing minimal, single-path representations demanded several nontrivial engineering solutions. For completeness, we summarize the most significant technical challenges and our corresponding contributions below:

\bheading{Building a Static JVM Emulator}
The JVM specification is designed for dynamic execution with concrete values, real memory allocation, and actual object instances. Emulating this behavior statically requires reimplementing the semantics of over 200 bytecode instructions, each with precise stack effects, type constraints, and control-flow implications. Every instruction must correctly model its impact on the operand stack, local variables, and heap state. A single incorrect stack effect causes all subsequent instructions to operate on wrong values. Beyond individual instructions, the emulator must handle the JVM's implicit behaviors: class loading sequences, field resolution, method dispatch, and exception propagation. Building a sound static model of this complexity is a substantial engineering undertaking. Therefore, our custom emulator contains more than $10\,000$ lines of code.

\bheading{State Splitting and Rewinding}
Unlike dynamic execution, which follows a single concrete path, our static emulation engine must explore all feasible paths within the given context. At each conditional branch, our engine must clone the entire execution state to explore both alternatives. This includes the call stack with program counters for each frame, local variables and operand stack per frame, static field values, object reference mappings, the recorded instruction trace, and branch-tracking metadata.

\bheading{Frame and Instance Tracking Without the JVM}
JVM implementations maintain call-stack frames and object instances in memory during execution. In purely static emulation, we must simulate these structures. Each frame maintains its own local variable array and operand stack, populated with symbolic objects rather than concrete data. The difficulty is that field reads and writes (\texttt{GETFIELD}/\texttt{PUTFIELD}) must resolve the correct object and field slot through our own tracking mechanism, and when values are unknown, placeholder objects must preserve stack shape without concrete semantics. Any mismatch in push/pop counts or local variable indices causes stack imbalances that invalidate the resulting bytecode.

\bheading{Adapting SWAT to Static Analysis}
SWAT was originally designed for dynamic symbolic execution with access to concrete runtime values. Repurposing it for purely static analysis required significant architectural changes to handle symbolic or unknown values where SWAT previously assumed concrete inputs.

\bheading{Loop and Recursion Bounding}
In static analysis, recursion and loops without concrete termination conditions can unroll indefinitely, resulting in non-termination. The challenge is detecting these situations reliably and bounding them without losing semantically important paths. 
Our engine therefore detects recursion. If so, the recursive trace terminates. The continuation logic beyond the recursive call is captured in the trace that followed the alternative branch at the recursion's origin. For intra-method loops, we detect back-edges and apply a single unrolling. Thus, a loop is either present or absent, resulting in two \TGs. However, once a loop is executed once, the conditional jump is copied into the resulting trace. Upon decompiling, the decompiler will detect the backreference and choose a suitable loop statement instead of leaving the statement as a conditional branch.

\bheading{Reconstructing Control-Flow Structure}
Java bytecode encodes control flow as unstructured conditional jumps rather than high-level if-else constructs. The challenge is that the compiler's output provides no explicit indication of whether a conditional has an else branch or simply falls through. To detect this, our analysis scans bytecode from conditional jump instructions to their targets, looking for indicators such as \texttt{GOTO} instructions that jump beyond the target  or \texttt{RETURN} instructions within the then-block. When no else is detected, the two exploration states are linked as ``neighbors'': if the then-branch completes normally, the engine skips the neighbor state since both paths converge to the same continuation, i.e., the one-sided branches from \Cref{sec:one-sided-branches}. However, if the then-branch ends in a \texttt{THROW}, the else-path must still be explored. Misidentifying these patterns causes either redundant exploration or missed paths.

\bheading{Resolving Interface and Virtual Calls}
Bytecode interface calls (\texttt{invokeinterface}) and virtual calls (\texttt{invokevirtual}) do not specify the concrete target method. Our engine resolves these dynamically-dispatched calls through two mechanisms. First, during class loading, we record which classes implement each interface and build the inheritance hierarchy. At call sites, the engine attempts to retrieve the receiver object's actual type from the symbolic operand stack by inspecting the stack position where the receiver resides  and extracting its tracked type. If the receiver is a placeholder or otherwise unresolved, the engine falls back to the recorded implementation list. If exactly one implementation exists, it uses that,  otherwise resolution fails. Once a candidate class is identified, the engine walks up the inheritance chain until it finds the actual method definition, correctly handling inherited methods.

\bheading{Modeling External Object Construction}
When the analyzed code receives objects from outside its scope (e.g., servlet request parameters), those objects lack visible construction history. The challenge is that we cannot simply skip initialization. Fields must have values for subsequent code to operate on, and constructors may establish invariants that later code depends upon. Therefore, we must synthesize the appropriate initialization sequences and handle constructor chaining to ensure that all values are correctly initialized.

\bheading{Pruning Infeasible Paths}
Exploring all branch combinations leads to exponential state explosion. The challenge is eliminating infeasible paths efficiently and maintaining soundness. We track lightweight variable values during emulation and eliminate branches that are logically unsatisfiable (e.g., a branch requiring a variable to be true when it was clearly set to false in the current trace).

\bheading{Static Initializer Ordering}
JVM static initializers (\texttt{<clinit>}) execute exactly once per class, triggered lazily on first use. The challenge in static emulation is that different exploration paths may encounter classes in different orders, yet each class must be initialized exactly once per trace to match JVM semantics. We must track which classes have been initialized within each trace, handle initialization dependencies between classes and ensure consistent states across cloned execution paths.

\bheading{Determining Analysis Scope}
Real-world JARs contain not only application code but also bundled dependencies. Analyzing everything would lead to a state explosion. However, analyzing too little misses relevant code paths. The challenge is automatically identifying the core-application code related to an entry point. Simple heuristics such as using the main class's package often fail for complex applications with multiple packages or unconventional structures. Our engine addresses this through density analysis: starting from the entry point, we perform many random walks along the call graph, counting how often each class is visited. We then score namespace prefixes by their visit frequency weighted by depth, selecting the namespace that best captures the application's core. This probabilistic approach adapts to diverse project structures without manual configuration.
Alternatively, users can explicitly specify the analysis scope as argument when the automatic detection is insufficient or when a specific subset of packages is desired.

\bheading{Producing Valid Bytecode from Traces}
After path exploration and slicing, reconstructing valid bytecode from the recorded instruction sequence is non-trivial. The challenge is that removing instructions during slicing disrupts the carefully balanced structure that the JVM verifier expects. The output must have balanced operand stacks at all points, consistent local variable indices, and correct jump target offsets despite removed instructions. Any inconsistency produces bytecode that fails verification or decompilation, providing no useful output.